\documentclass{jfm}
\pdfoutput=1

\usepackage[pdftex]{graphicx}
\usepackage{natbib}
\usepackage[pdftex]{graphicx}
\usepackage{bm}
\usepackage{amssymb}
\usepackage{amsbsy}
\usepackage{color}
\usepackage{mathrsfs,mathcomp}
\usepackage{amsmath}
\usepackage{amsfonts}
\usepackage{amssymb}
\usepackage{amsbsy}
\usepackage{color}
\usepackage{mathrsfs}
\usepackage{graphicx}
\usepackage{subfigure}

\newcommand{\we}{{\it We}}

\title[Drop deformation by laser-pulse impact]{Drop deformation by laser-pulse impact}% Drop deformation and propulsion by laser pulse impact
\author[H. Gelderblom et al.]
{H\ls A\ls N\ls N\ls E\ls K\ls E\ns G\ls E\ls L\ls D\ls E\ls R\ls B\ls L\ls O\ls M$^1$,\ns H\ls E\ls N\ls R\ls I\ns L\ls H\ls U\ls I\ls S\ls S\ls I\ls E\ls R$^2$,\ns \break
A\ls L\ls E\ls X\ls A\ls N\ls D\ls E\ls R\ns L.\ns K\ls L\ls E\ls I\ls N$^1$,\ns
	W\ls I\ls L\ls C\ls O\ns  B\ls O\ls U\ls W\ls H\ls U\ls I\ls S$^1$,\ns \break 
	D\ls E\ls T\ls L\ls E\ls F\ns L\ls O\ls H\ls S\ls E$^1$,\ns  E\ls M\ls M\ls A\ls N\ls U\ls E\ls L\ns V\ls I\ls L\ls L\ls E\ls R\ls M\ls A\ls U\ls X$^3$\ns \break
 \and{} J\ls A\ls C\ls C\ls O\ns H.\ns S\ls N\ls O\ls E\ls I\ls J\ls E\ls R$^{1,4}$}

\affiliation{$^1$Physics of Fluids Group, Faculty of Science \& Technology,  J.M.\ Burgers Center for Fluid
Dynamics, University of Twente, P.O.\ Box 217, 7500 AE Enschede, The Netherlands\\[\affilskip]
$^2$ IUSTI UMR 7343, CNRS $\&$ Aix-Marseille Universit\'e, 13453 Marseille, France\\[\affilskip]
$^3$ IRPHE, Aix-Marseille Universit\'e, 13384 Marseille CEDEX 13, France\\[\affilskip]
$^4$ Mesoscopic Transport Phenomena, Eindhoven University of Technology, Den Dolech 2, 5612 AZ Eindhoven, The Netherlands
}
\pubyear{??}
\volume{??}
\pagerange{??--??}
\date{?? and in revised form ??}
\begin{document}

\maketitle

\begin{abstract}
	A free-falling absorbing liquid drop hit by a nanosecond laser-pulse experiences a strong recoil-pressure kick. As a consequence, the drop propels forward and deforms into a thin sheet which eventually fragments. We study how the drop deformation depends on the pulse shape and drop properties. We first derive the velocity field inside the drop on the timescale of the pressure pulse, when the drop is still spherical. This yields the kinetic-energy partition inside the drop, which precisely measures the deformation rate with respect to the propulsion rate, before surface tension comes into play. On the timescale where surface tension is important the drop has evolved into a thin sheet. Its expansion dynamics is described with a slender-slope model, which uses the impulsive energy-partition as an initial condition. Completed with boundary integral simulations, this two-stage model explains the entire drop dynamics and its dependance on the pulse shape: for a given propulsion, a tightly focused pulse results in a thin curved sheet which maximizes the lateral expansion, while a uniform illumination yields a smaller expansion but a flat symmetric sheet, in good agreement with experimental observations.
\end{abstract}

\section{Introduction}\label{intro}

\begin{figure}\begin{center}
	\includegraphics[width=\linewidth]{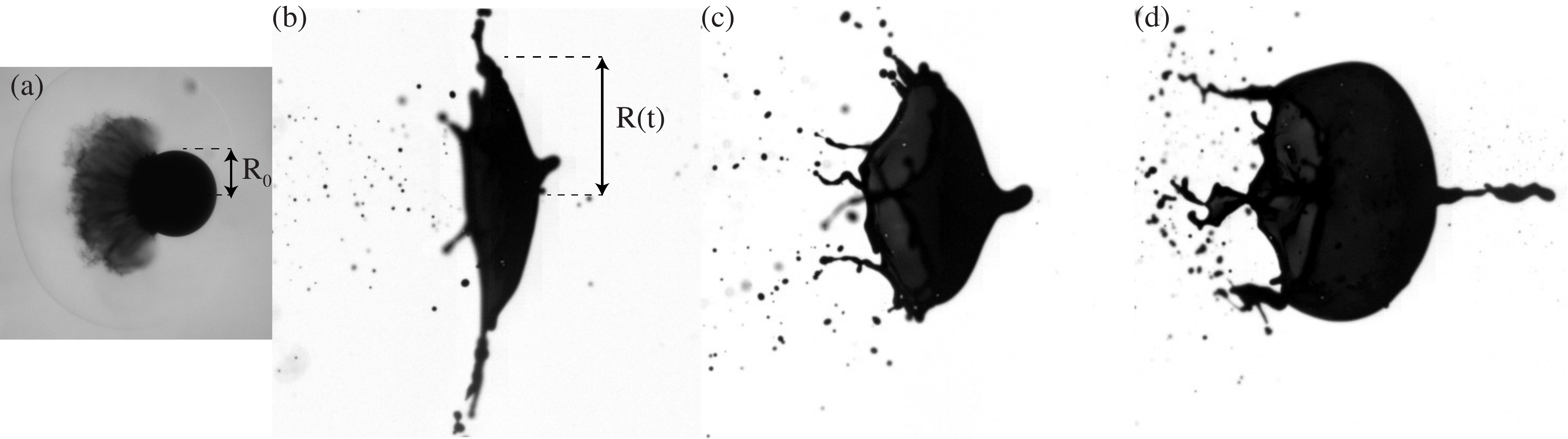}
	\caption{Response of a free-falling dyed water drop of initial radius $R_0=0.9$\,mm to a 10\,ns laser pulse ($\lambda=532$\,nm) impacting from the left. (a) $t=5\,\mu$s after the pulse, a shock wave has propagated in the air and a mist cloud has developed at the drop surface, but the drop itself has not yet moved. (b-d) $t=1.2$\,ms after the pulse, the drop has propelled and deformed into a thin sheet (same magnification as in (a)), whose shape and lateral expansion $R(t)$ depend not only on the energy $E$ absorbed by the drop, but also on the width of the laser beam on the drop surface $\sim\sigma R_0$ (see \S \ref{results}): (b) uniform illumination of the drop ($\sigma\simeq 0.75$, $E=29$\,mJ), (c) slightly focused laser ($\sigma\simeq0.48$, $E=20$\,mJ), and (d) tightly focused laser ($\sigma\simeq0.29$, $E=20$\,mJ).}\label{fig1}
\end{center}\end{figure}

	A laser pulse interacting with an absorbing liquid body can deposit a finite amount of energy, concentrated both in time and space, which eventually triggers a dramatic hydrodynamic response. Focused nanosecond pulses have for instance been used to induce cavitation in liquids confined in capillary tubes \citep{Vogel:1996, Sun:2009, Tagawa:2012}, or jetting and spraying in sessile drops \citep{Thoroddsen:2009}. These situations involving a liquid close to a wall result in localized flows. By contrast, we consider here the situation of a mobile liquid body: the impact of a nanosecond laser pulse onto an absorbing unconfined liquid drop, which, as first described by \cite{Klein:2015}, has a global hydrodynamic response to the pulse: the drop propels forward at a speed of several meters per second, strongly deforms and eventually fragments (see Fig.\ \ref{fig1}). This dynamics is similar to that following a mechanical impact such as on a solid substrate or a pillar, which has been studied thoroughly \citep[see e.g.][]{Clanet:2004, Yarin:2006, Villermaux:2011, Mandre:2012, Riboux:2014, Thoroddsen:2016}, including a few studies on the fragmentation of the drop \citep{Villermaux:2007, Xu:2007, Villermaux:2009, Villermaux:2011, Riboux:2014}. A laser proves to be an adequate tool to vary the extension of the impact without affecting the initial drop geometry. However, how a drop deforms and fragments as a results of a laser impact are still a largely open questions.

An important application of drop deformation by laser-pulse impact
is found in laser-produced plasma light-sources for extreme ultraviolet (EUV) nanolithography. In these sources a nanosecond laser pulse pre-shapes a falling liquid tin drop into a thin sheet, which is subsequently ionized by a second laser pulse \citep{Mizoguchi:2010, Banine:2011}. To maximize the conversion of liquid tin to plasma a precise control of the drop shape and stability that result from the first laser impact is crucial. That is, the dynamic response of a liquid drop to the impact of a laser pulse has to be resolved.
	
	In a previous study \citep{Klein:2015} we focussed on the question of how the laser transfers momentum to the liquid body. We showed that the key driving mechanism for the drop propulsion and deformation observed in experiments is the local and asymmetric boiling of the liquid induced by the absorption of the laser energy on the illuminated side of the drop. In a dyed (and hence absorbing) drop this absorption occurs in a thin, superficial layer of liquid, whose thickness is set by the penetration depth of the laser. As a result this layer boils and a shock wave is emitted in the surrounding air, followed by the directive emission of vapour and mist (see Fig.\ \ref{fig1}a). This vaporization applies a recoil pressure on the drop surface which both deforms the drop and propels it forward (Fig.\ \ref{fig1}b-d) at a velocity
\begin{equation}
	U\sim \frac{E-E_{\rm th}}{\rho R_0^3 \Delta H}u.\label{propspeed}
\end{equation}
This propulsion velocity scales linearly with the absorbed laser energy $E$ beyond the threshold energy $E_{\rm th}$ needed to heat the liquid layer to the boiling point, where $\rho$ is the liquid density, $R_0$ is the initial drop radius, $\Delta H$ the latent heat of vaporisation and $u$ the thermal speed of the expelled vapour. The drop propulsion is accompanied by a lateral expansion that scales as 
\begin{equation}
	\frac{R_{\rm max}-R_0}{R_0}\sim \we^{1/2},\label{expanKlein}
\end{equation}
where the Weber number is defined as $\we=\rho R_0 U^2/\gamma$, and $\gamma$ is the liquid surface tension. Hence, both the propulsion speed and the maximal radius of expansion are proportional to the laser pulse energy (beyond the threshold). However, not only the energy of the laser pulse, but also the pulse shape and focus have a strong influence on the drop deformation and propulsion, as Fig.\ \ref{fig1}b-d shows. Although the absorbed laser energy is similar in the three cases shown, the resulting drop shapes differ completely: an unfocussed laser beam deforms the drop into an almost flat sheet, whereas a focussed beam gives rise to a strongly curved, bag-like drop shape. 

\begin{figure}\begin{center}
	\includegraphics[width=.8\textwidth]{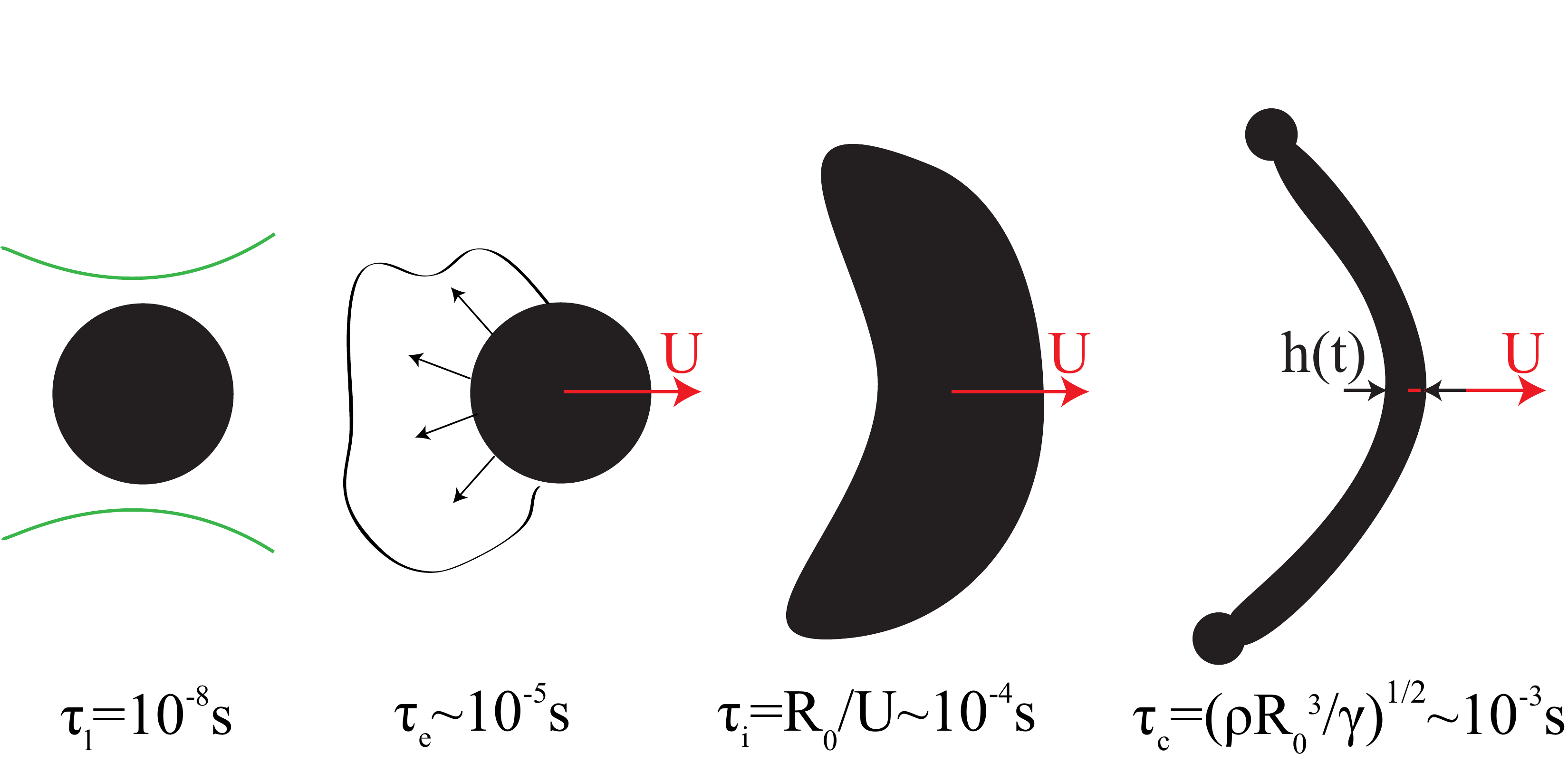}
	\caption{Illustration of the timescales separation in the problem. The laser interacts with the drop on time $\tau_\ell$, the drop reaches its centre-of-mass velocity $U$ on the vapour expulsion time $\tau_{\rm e}$. The drop subsequently deforms on the inertial time $\tau_{\rm i}$ into a thin sheet with time-dependent thickness $h(t)$, which undergoes a surface-tension limited expansion on the capillary time $\tau_{\rm c}$.} \label{fig2}
\end{center}\end{figure}

	 Before seeking for understanding these differences it is worth remembering the clear separation of the timescales involved in the problem \citep{Klein:2015}, which we illustrate in Fig.\ \ref{fig2}. The effect of a few milli-joules laser pulse with a duration $\tau_\ell\sim 10^{-8}$\,s onto a liquid drop can successfully be modeled as a recoil-pressure pulse exerted on the drop surface for a duration $\tau_{\rm e}\sim 10^{-5}$\,s, the typical timescale for the vapour and mist ejection \citep{Klein:2015}. It is clear from Fig.\ \ref{fig1}a, that on this timescale the drop does not deform: both the laser pulse duration $\tau_\ell$ and the vapour-recoil duration $\tau_{\rm e}$ are much shorter than the inertial and capillary timescales, respectively $\tau_{\rm i}=R_0/U\sim 10^{-4}$\,s and $\tau_{\rm c}=\sqrt{\rho R_0^3/\gamma}\sim10^{-3}$\,s, on which the drop propels, deforms and fragments (Fig.\ \ref{fig1}b-d).
 
 	The present work aims to elucidate how the laser-pulse shape and focus affect the drop deformation and propulsion. To this end, we employ both analytical and numerical modeling and make use of the separation of the timescales $\tau_\ell\ll \tau_{\rm e}\ll \tau_{\rm i}<\tau_{\rm c}$. In \S 2 and \S 3 we follow a pressure impulse approach as described by \citet[\S 6.10]{Batchelor} and \citet{Antkowiak:2007} to obtain, for an arbitrary pulse shape, the velocity field in the drop and the kinetic-energy partition between the deformation and the translation of the drop on the timescale $\tau_{\rm e}$, i.e.~the initial lateral expansion rate of the drop relative to its propulsion speed. Surprisingly, we find that the maximum expansion rate is achieved when one focusses the laser pulse into a tight spot, whereas a flat (symmetric) expanding drop is obtained only with a uniform laser-beam profile. On the intermediate timescale $\tau_{\rm i}$ the drop deforms significantly and a purely ballistic approach is no longer applicable. We use in \S3 a numerical boundary integral (BI) method \citep{Oguz:1993, Power:1995, Bergmann:2009, Gekle:2010, Bouwhuis:2012} to confirm the main features of the deformation and the precise detail of the flow. 
For an unfocussed laser pulse (Fig.~\ref{fig1}b) the drop evolves into a flat, thin sheet. In \S 4 we use the kinetic-energy partition obtained from
the early-time analytical model and follow the method of \citet{Villermaux:2009} to describe the surface-tension limited expansion of this sheet on the late timescale $\tau_{\rm c}$.

\section{Problem formulation \& methods}

	We consider the response of a liquid drop to a pressure pulse, i.e.~a pressure field with magnitude $p_{\rm e}$ applied at the interface on one side of the drop for a duration $\tau_{\rm e}$. The absolute impulse scale $p_{\rm e}\tau_{\rm e}$ sets the propulsion velocity of the drop through momentum conservation (see (\ref{momcon}) below). As we discussed above, this velocity is in turn directly related to the laser pulse energy through (\ref{propspeed}). The problem thus amounts to determining the shape and the rate of deformation of the drop. In \S \ref{earlymodel} we introduce an analytical model for the early-time dynamics of the drop ($t\sim \tau_{\rm e}$). The BI model used to simulate the drop dynamics at later times ($t\sim \tau_{\rm i}, \tau_{\rm c}$) is discussed in \S \ref{BI}.
\begin{figure}\begin{center}
	\includegraphics[width=0.5\linewidth]{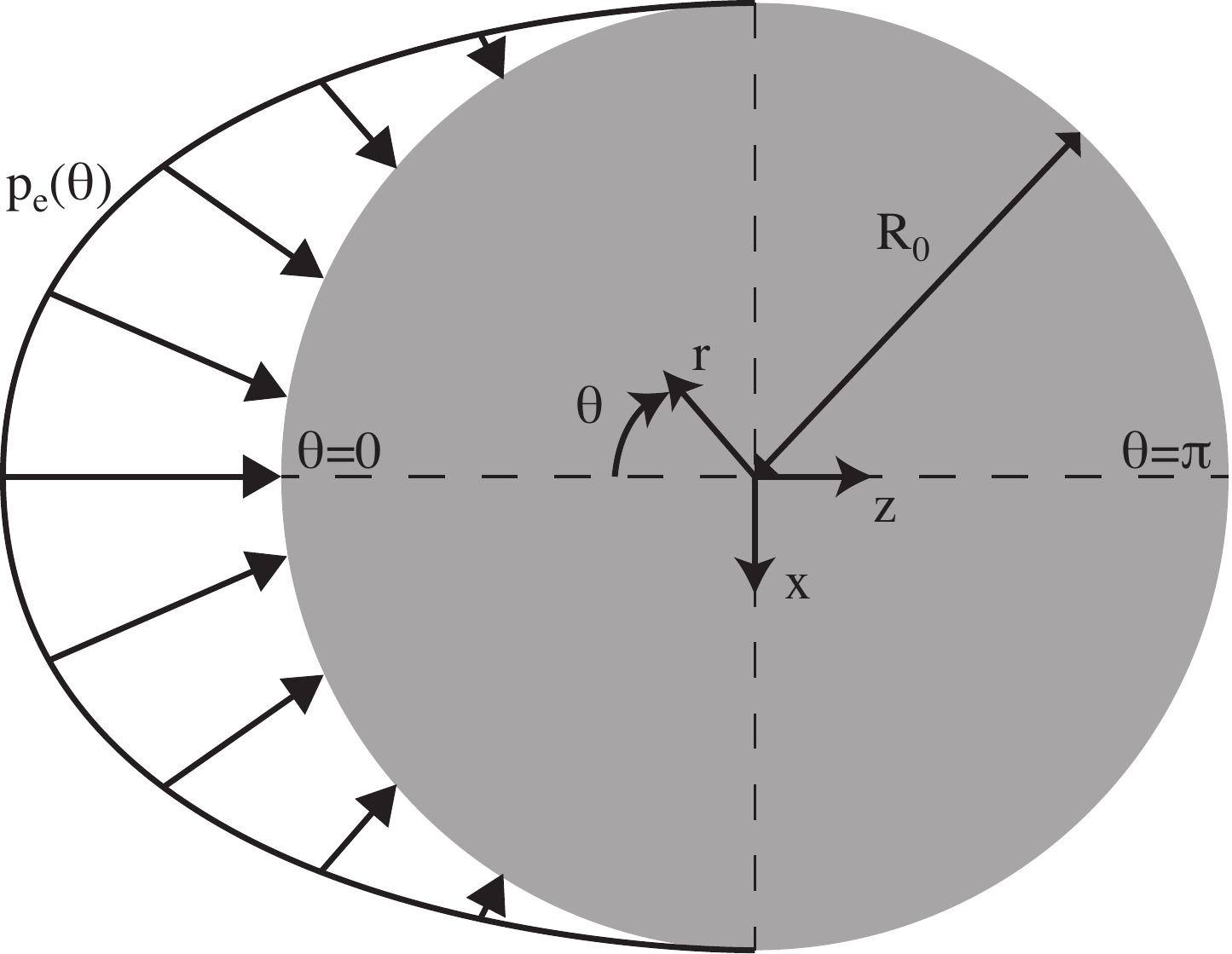}
	\caption{Sketch of the problem. The axisymmetric pressure pulse $p_{e}(\theta)$ applies on the  surface of a drop of radius $R_0$. The spherical ($r,\theta,\phi$) and cartesian ($x,y,z$) coordinates systems are indicated.}\label{fig3}
\end{center}\end{figure}

\subsection{Early time dynamics: analytical model}\label{earlymodel}
We characterise the ratio between the inertial timescale on which the drop deforms and the vapour-expulsion time on which the drop acquires it centre-of-mass speed by the impact number
\begin{equation}
I=\frac{\tau_i}{\tau_e}=\frac{R}{U\tau_e}.\label{impactno}
\end{equation}
Note that since $I\gg 1$ the drop does not deform on the time-scale of the pressure pulse, as is shown in Fig.~\ref{fig1}a. To find the post-impact velocity field we therefore naturally consider the impulsive response of a spherical drop. Figure \ref{fig3} shows a sketch of the problem geometry and indicates both the spherical $(r,\theta,\phi)$ and cartesian coordinates $(x,y,z)$. Both the initial configuration and the pressure pulse are symmetric around the laser axis ($z$-axis), and we therefore seek a velocity field that is symmetric too. The pressure pulse applied on the drop surface sets the fluid in motion inside the entire drop. The axial propulsion speed $U$ of the drop (see Fig.\ \ref{fig3}), i.e.\ its centre-of-mass velocity, follows from the global momentum conservation
\begin{equation}\label{momcon}
	\int_0^{\tau_{\rm e}}\int_A p_{\rm e}\mathbf{e_z}\cdot\mathrm{d}\mathbf{A}\mathrm{d}t=\frac{4}{3}\pi\rho R_0^3 U,
\end{equation}
with $A$ the surface of the drop.

	To describe the flow field inside the drop we follow the same approach as \citet[\S 6.10]{Batchelor} and \cite{Antkowiak:2007}. The pressure field establishes on the sonic timescale $R_0/c\sim 10^{-6}$\,s, with $c$ the speed of sound in the liquid, which is much shorter than the pressure pulse duration $\tau_{\rm e}\sim 10^{-5}$\,s. Hence, on time $\tau_{\rm e}$ the pressure field is well established. As the Reynolds number in these experiments is typically large (Re $\sim 10^3$) the flow is inviscid. Since moreover $I\gg 1$ (i.e.\ $\partial \mathbf{u}/\partial t\gg (\mathbf{u}\cdot \nabla)\mathbf{u}$), the impulsive acceleration of the drop during the pulse follows
\begin{equation}
	\frac{\partial\mathbf{u}}{\partial t}\approx-\frac{1}{\rho}\nabla p,\label{ns1}
\end{equation}
with $\mathbf{u}(r,\theta,\phi)$ the velocity and $p$ the pressure inside the drop relative to the ambient pressure. Incompressibility ($U\ll c$) implies, upon taking the divergence of (\ref{ns1}), that the pressure field is harmonic:
\begin{equation}
\Delta p=0. \label{ns2}
\end{equation}
The velocity field just after the pressure pulse is then obtained by integration of (\ref{ns1}) over time
\begin{equation}
	\mathbf{u}\approx-\frac{1}{\rho}\nabla \int_0^{\tau_{\rm e}} p(\tau)\mathrm{d}\tau= -\frac{\tau_{\rm e}}{\rho}\nabla p,\quad \tau_{\rm e} \leq t \ll \tau_{\rm i}, \label{velfield}
\end{equation}
where $p_{\rm e}$ refers to the time-averaged recoil pressure exerted on the drop during the pulse. From momentum conservation (\ref{momcon}) it follows that the drop speed $U$ scales as
\begin{equation}
	U\sim \frac{p_{\rm e}\tau_{\rm e}}{\rho R_0}.\label{velscale}
\end{equation}
From now on, we use the scaled time $t/\tau_{\rm e}$, radial coordinate $r/R_0$, pressure $p/p_{\rm e}$, and velocity $\rho R_0 \mathbf{u}/p_{\rm e}\tau_{\rm e}$.

	The shape of the pressure pulse $f(\theta)$ arises as the boundary condition on the drop surface
\begin{equation}
	p(r=1,\theta)=f(\theta),\label{presbc}
\end{equation}
which we normalize such that the axial momentum is equal to one, i.e.
\begin{equation}
	\frac{4}{3}\pi U=\int_A f(\theta) \mathbf{e}_z\cdot\mathrm{d}\mathbf{A}=2\pi\int_0^\pi f(\theta) \cos\theta\sin\theta\mathrm{d}\theta=1.\label{zmom}
\end{equation}
This choice sets the (dimensionless) centre-of-mass velocity of the drop $U=3/(4\pi)$ and the associated translation kinetic energy 
\begin{equation}
	E_{\rm k,cm}=\frac{2}{3}\pi U^2=\frac{3}{8\pi},
\end{equation}
independently of the choice of $f(\theta)$.

	To solve the Laplace equation (\ref{ns2}) in spherical coordinates we decompose the pressure field into Legendre polynomials $P_\ell$ 
\begin{equation}
	p(r,\theta)=\sum_{\ell=0}^{\infty} A_\ell r^\ell P_\ell\left(\cos\theta\right),\label{sol1}
\end{equation}
which coefficients
\begin{equation}
	A_\ell=\frac{2\ell+1}{2}\int_0^\pi f(\theta)P_\ell\left(\cos\theta\right)\sin\theta\mathrm{d}\theta,\label{coef1}
\end{equation}
are obtained by the projection of the boundary condition (\ref{presbc}). From (\ref{zmom}) one observes that $A_1=U$.

	The solution (\ref{sol1}-\ref{coef1}) can now be used to describe the drop response to any pressure- and hence any laser-beam profile. The corresponding velocity field is computed from (\ref{velfield}). While by convention $E_{\rm k,cm}$ does not depend on the pressure-pulse shape, the total amount of kinetic energy that has to be put into the drop to acquire this propulsion does. It is given by
\begin{equation}
	E_{\rm k}=\frac{1}{2}\int_V \mathbf{u}^2\mathrm{d}V=\pi\int_0^1\int_0^\pi \left(u_r^2+u_\theta^2\right)r^2\sin\theta\mathrm{d}\theta\mathrm{d}r,\label{etot}
\end{equation}
with $V$ the drop volume. As we will see in \S \ref{gauss}, it is convenient to define the partition
\begin{equation}
	\frac{E_{\rm k,d}}{E_{\rm k}}=1-\frac{E_{\rm k,cm}}{E_{\rm k}}\label{energy}
\end{equation}
between the deformation kinetic energy of the drop $E_{\rm k,d}$ (i.e. the kinetic energy remaining in the co-moving frame) and the total kinetic energy (\ref{etot}).% depends on the shape of the pressure pulse that is applied to the drop.

\subsection{Boundary integral simulations}\label{BI}

	The analysis above applies when the drop shape does not deviate too much from a sphere ($t\sim \tau_{\rm e}\ll \tau_{\rm i}$). To obtain the details of the subsequent drop-shape evolution one needs to solve the axisymmetric potential flow problem in the deforming shape. To this end, we employ the boundary integral (BI) method described by \cite{Bergmann:2009, Gekle:2010}, which has already been successfully used to study drop deformation during mechanical impact \citep{Bouwhuis:2012}, as well as that due to a laser impact \citep{Klein:2015}. BI is a powerful method to study the drop dynamics at later times $t\sim \tau_{\rm i}$, when the drop shape changes significantly.

\section{Results}\label{results}

	We will now use the analytical model and BI simulations to explore the role of the laser-pulse shape, i.e.~of the pressure-pulse shape, on the deformation of the drop. Indeed, the pressure boundary condition (\ref{presbc}) introduced above is the actual pressure on the drop surface, which is typically proportional to the local laser fluence weighted by the cosine of the incident angle of the incoming rays on the drop surface. Typical laser-beam profiles used in experiments have a Gaussian or flat-top (uniform) shape. We consider a Gaussian pulse with a finite arbitrary width in \S \ref{gauss}, the limits of a perfectly focussed beam in \S\ref{delta} and that of a uniform laser-beam profile, i.e.\ a cosine pressure pulse applied on one side of the drop, in \S \ref{flat}.

\subsection{Gaussian laser-beam profile}\label{gauss}
	For simplicity, we first consider a pressure pulse that applies over the entire drop surface. The effect of restricting the interaction to the side that is actually illuminated by the laser will be discussed in \S \ref{flat}. Since our aim is to understand the influence of the laser focus on the drop-shape evolution we also neglect the angular dependence $\cos{\theta}$ of the pressure profile. The Gaussian-shaped pressure boundary condition (\ref{presbc}) then reads
\begin{equation}
	f(\theta)=c \exp\left[-\theta^2/(2 \sigma^2)\right],\label{bc}
\end{equation}
where $\sigma$ is a measure for the width of the pulse and the prefactor
\begin{equation}
	c=\frac{2 \sqrt{2}}{\sigma \pi^{
 3/2}\exp\left[-2 \sigma^2\right]  \left(2 \mathrm{Erfi}\left[\sqrt{2} \sigma\right]- \mathrm{Erfi}\left[\frac{i\pi + 2 \sigma^2}{\sqrt{2} \sigma}\right]- 
   \mathrm{Erfi}\left[\frac{-i\pi + 2 \sigma^2}{\sqrt{2} \sigma}\right] 
   \right)}.
\end{equation}
ensures the normalization (\ref{zmom}). The resulting coefficients (\ref{coef1}) are calculated by numerical integration. The convergence of series (\ref{sol1}) depends on the value of $\sigma$, but in general 20 terms are sufficient to obtain accurate results (except in the limit $\sigma\to 0$, which has to be treated separately and will be discussed in \S \ref{delta}).

\subsubsection{Global features}

\begin{figure}\begin{center}
	\includegraphics[width=\linewidth]{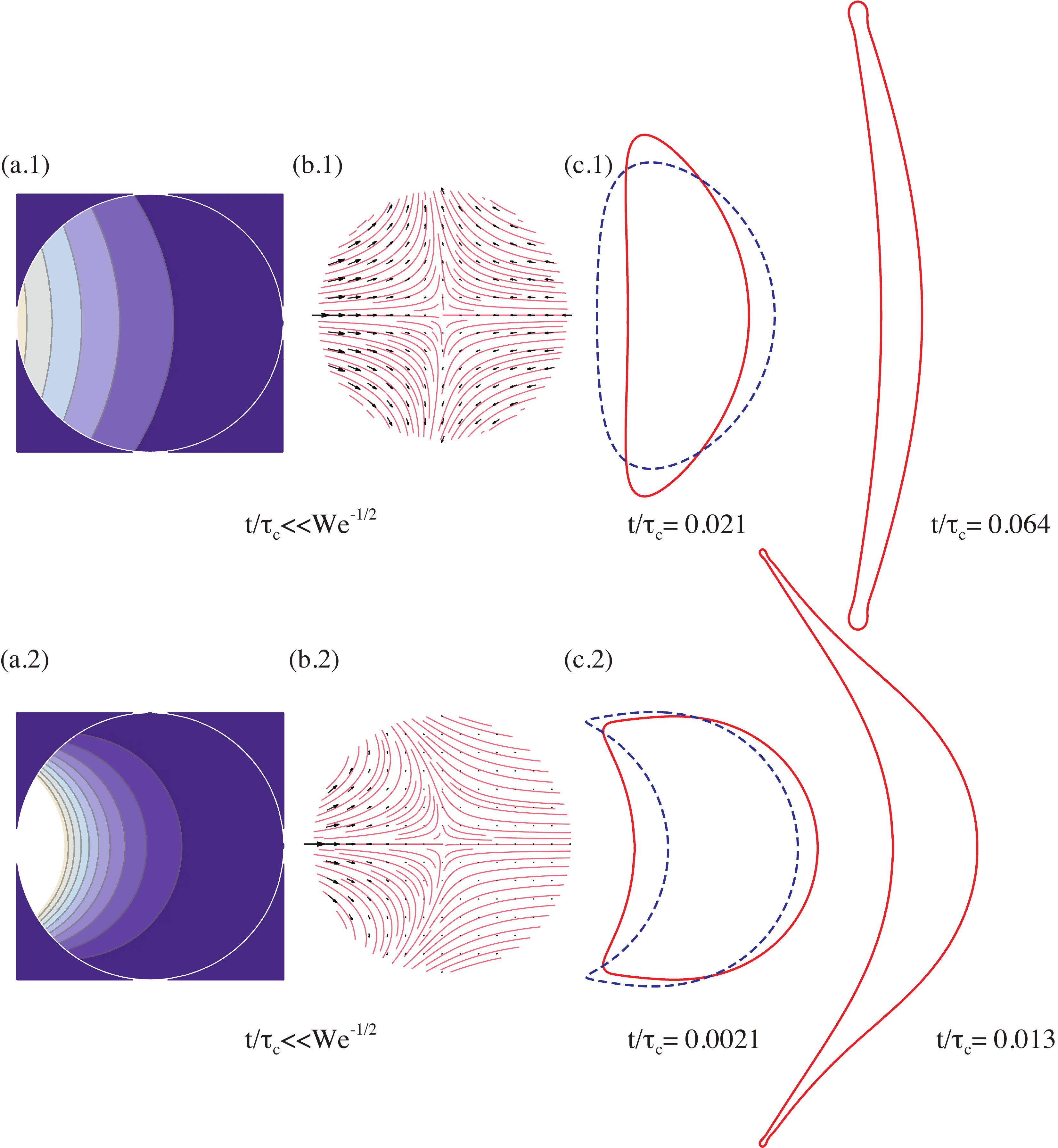}
	\caption{(a) Iso-pressure lines inside the drop at early times ($t/\tau_{\rm c}\ll \we^{-1/2}$) for Gaussian pressure pulses with $\sigma=\pi/4$ (a.1) and $\sigma=\pi/8$ (a.2). (b) Streamlines of the early-time velocity field ($t\sim\tau_{\rm e}$) in the co-moving frame (drop centre-of-mass velocity subtracted). (c) Drop contours illustrating the evolution of the deformation in the analytical model (blue dashed lines) obtained by advecting the material points on the drop surface (see text) and in the BI simulations (red solid lines drawn on the same scale as in (a-b), $\we=790$). Note that the  expansion is much faster for $\sigma=\pi/8$ (c.2) than for $\sigma=\pi/4$ (c.1), the contours being represented earlier in the latter case.}\label{fig4}
\end{center}\end{figure}

	We explore the effect of the focussing of the laser beam on the drop deformation by varying the pulse width $\sigma$, thereby mimicking the situation shown in Fig.\ \ref{fig1}b-d. In Fig.\ \ref{fig4} we show a plot of the resulting pressure and velocity fields inside the drop for a uniform pressure pulse ($\sigma=\pi/4$) and a more focussed one ($\sigma=\pi/8$). In these (and the following) plots, the series solution (\ref{sol1}) is cut after 20 terms. The velocity fields shown in Fig.\ \ref{fig4} are in the co-moving frame: we subtracted the centre-of-mass velocity of the drop to clearly illustrate the deformation of the drop during its translational motion. The analytical solution (\ref{velfield},\ref{sol1}) is strictly valid only as long as the domain is spherical. However, we can obtain a first-order approximation of the deformed drop shape shortly after the pressure kick by advecting the material points on the drop surface. The drop surface at time $t$ is then given by $\mathbf{r_d}(\theta,t)=\mathbf{e_r}+\left[u_r(1,\theta)\mathbf{e_r}+u_\theta(1,\theta)\mathbf{e_\theta}\right]It$, with $I$ given by (\ref{impactno}); see Fig.\ \ref{fig4}c (blue dashed lines). This mere extrapolation must of course only be considered for qualitative and illustrative purposes. For a quantitative prediction, one needs to consider hydrodynamic interaction and solve for the pressure (\ref{ns2}) in the deformed drop, which is done in the BI simulations. A few drop contours obtained from this simulation for a Weber number of 790 are shown in Fig.\ \ref{fig4}c (red solid lines). 

	From Fig.\ \ref{fig4} we observe that an unfocussed pulse leads to a velocity field that is almost symmetric around the vertical mid-plane (Fig.\ \ref{fig4}b.1). As a consequence, the eventual drop shape that will result from this pressure pulse is almost symmetric and flat, as is indeed observed in the BI results in Fig.\ \ref{fig4}c.1 and (to some extend) in our experimental results in Fig.\ \ref{fig1}b. By contrast, a focussed pulse naturally leads to more curved iso-pressure lines and the eventual drop will also be more curved (Fig.\ \ref{fig4}c.2), which agrees with our experimental observations in Fig.\ \ref{fig1}c-d. The BI results show that at later times ($t>\tau_{\rm i}$ and hence $t/\tau_{\rm c} > \we^{-1/2}$), the drop deforms into a thin sheet bordered by a rim. For the unfocussed pulse ($\sigma=\pi/4$, Fig.\ \ref{fig4}c.1) this sheet is relatively flat and has an approximately uniform thickness, except for the rim itself. For the focussed pulse ($\sigma=\pi/8$, Fig.\ \ref{fig4}c.2) the resulting sheet has a stronger curvature with a clearly non-uniform thickness, and the expansion is much faster than for the focussed pulse (note the difference in timescales between Fig.\ \ref{fig4}c.1 and c.2).

	In the BI simulations, the recession of the sheet edge eventually leads to the formation of undamped surface waves and a Bernoulli suction that results into the successive detachments of liquid rings from the edge. This pinch-off is an artefact of the simulation caused by the lack of viscous damping and the assumption of axial symmetry, as discussed by \citet{Peters:2013}, and is clearly irrelevant to the physical fragmentation processes that actually occur. This artefact however has a negligible influence on the early-time expansion and evolution of the sheet thickness away from the rim. We therefore use the simulations until the first pinch-off event occurs.

\subsubsection{Kinetic-energy partition: deformation versus translation}

\begin{figure}\begin{center}
	\includegraphics[width=0.7\linewidth]{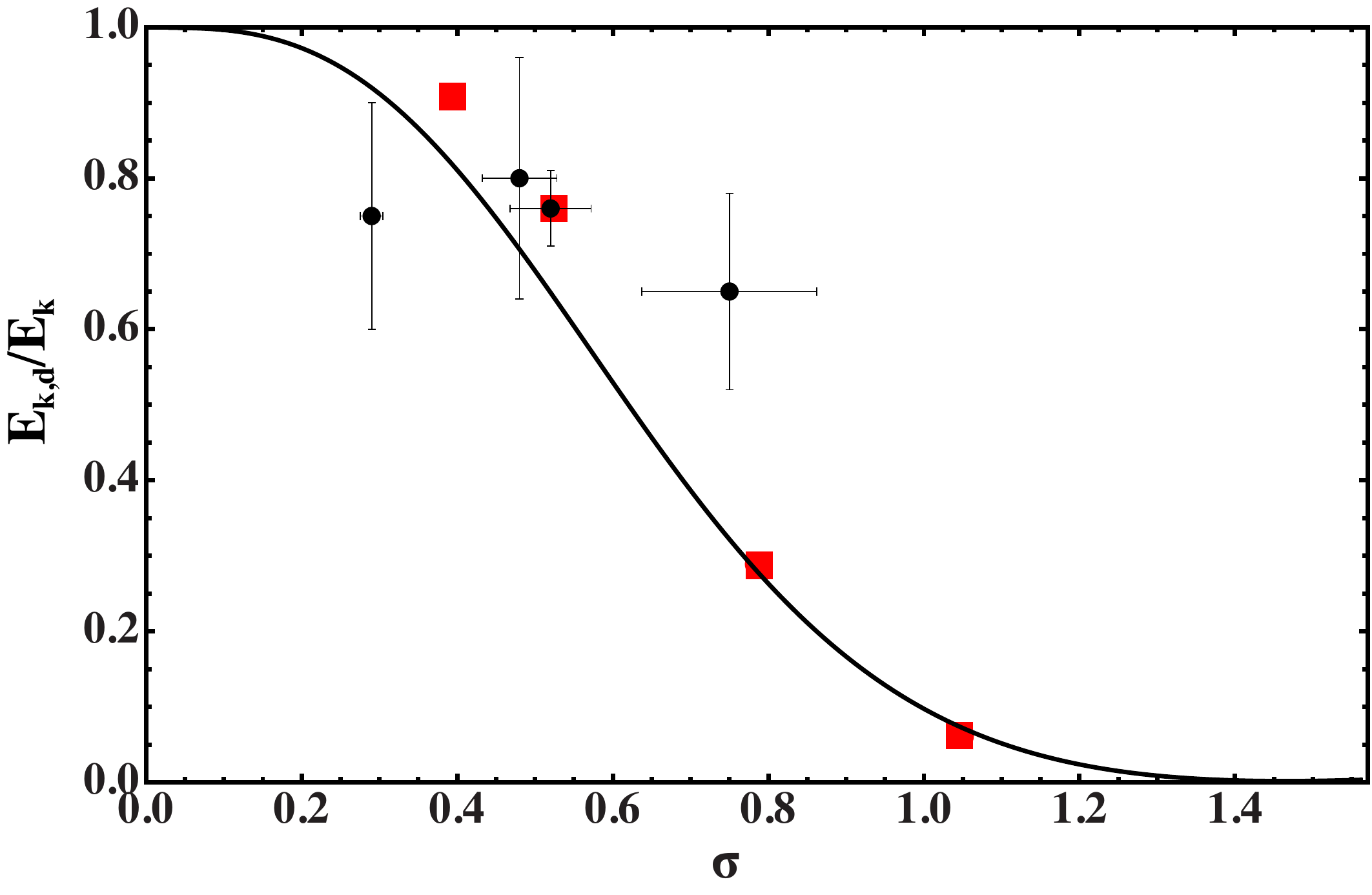}
	\caption{Kinetic-energy partition (\ref{energy}) as a function of the pulse width $\sigma$. For a tightly focused beam (small $\sigma$) almost all the energy goes into deforming the drop without translating it. For an unfocussed pulse the drop only translates but hardly deforms. The black dots are experimental data obtained, by a method described in Appendix \ref{app2}, from the three events shown in Fig.\ \ref{fig1}b-d and from the experiments in \citet{Klein:2015} (four realisations of $\sigma=\pi/6\simeq 0.52$). The red squares are obtained from BI simulations using the same method to estimate the energy partition as in the experiments (see Appendix \ref{app2}).}\label{fig5}
\end{center}\end{figure}

	We now use the analytical results (\ref{velfield},\ref{sol1}) to quantify the effect of focussing the laser on the expansion rate of the drop relative to its propulsion velocity. Figure \ref{fig5} shows the kinetic-energy partition (deformation to total kinetic-energy ratio) (\ref{energy}) as a function of the pulse width $\sigma$. We also plot estimates for  the energy partition obtained from the three experimental cases shown in Fig.\ \ref{fig1}b-d and from the data of \citet{Klein:2015} (black circles). In Appendix \ref{app2} we explain in detail the (non-trivial) steps that are taken to obtain these estimates from the experimental data. For comparison, we applied the same method to the BI simulations (red squares), which confirms the validity of our method (see Appendix \ref{app2} for further discussion). Given the uncertainties in the experimental estimates in particular for the focussed laser pulse, as discussed in Appendix \ref{app2}, we cannot expect a quantitative agreement with theory. However, Fig.\ \ref{fig5} shows that the experimental data points qualitatively confirm the theoretical prediction: the more focussed the laser pulse, the more energy is used to deform the drop rather than to translate it.

	Figure \ref{fig5} shows that for a tightly focussed beam (small $\sigma$) almost all the energy goes into deforming the drop and hardly any into translating it: the energy ratio $E_{\rm k,d}/E_{\rm k}\to 1$ as $\sigma\to 0$. Indeed, the total kinetic energy required to maintain a constant centre-of-mass speed diverges as the pressure pulse becomes more localized. We will discuss the limiting case when the pressure pulse comes down to a Dirac-delta pulse in more detail in \S \ref{delta}. By contrast, when the pulse is very broad (large $\sigma$) the drop experiences a pressure from all sides, such that it does not deform but only translates and $E_{\rm k,d}/E_{\rm k}\to 0$. Note again that $\sigma\ll1$ does not represent a large directional laser beam applying only on one side of the drop, which will be considered below, but rather an isotropic illumination of the drop. In fact, the Gaussian pressure pulse that is the most relevant to a uniform laser-beam profile (see \S \ref{flat}) has $\sigma\simeq 0.73$, which is fairly unidirectional and close to the $f(\theta) \propto \cos\theta$ profile due to the local incidence of the laser on the curved drop surface.

\begin{figure}\begin{center}
	\includegraphics[width=\textwidth]{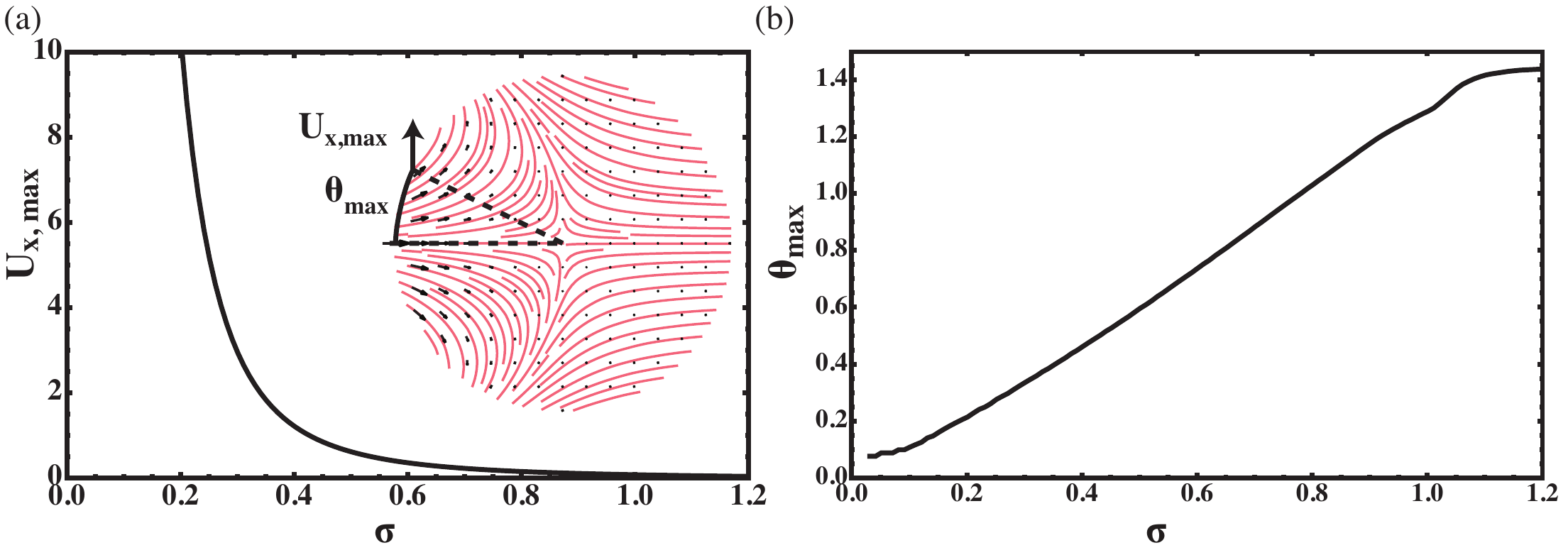}
	\caption{(a) Maximal expansion velocity $U_{x,{\rm max}}$ (in the $x$-direction) along the drop surface as a function of $\sigma$. The more focussed the pulse, the faster the drop expands. The inset shows the velocity field in the co-moving frame for $\sigma=\pi/8$, and sketches $U_{x,{\rm max}}$ and its angular location $\theta_{\rm max}$. (b) $\theta_{\rm max}$ as a function of $\sigma$. For a focussed pulse the maximal expansion velocity is observed around $\theta=0$, i.e. on the pulse axis of symmetry. No data is shown for $\sigma\to 0$ and $\sigma\to\pi/2$ since in these limits the series (\ref{sol1}) does not converge or the deformation velocity becomes negligible, respectively.}\label{fig6}
\end{center}\end{figure}

	Figure \ref{fig5} shows that a focussed pressure pulse leads to a stronger drop deformation. This does not necessarily mean that the drop will also experience a larger lateral expansion, since the energy could be used to deform the drop into a strongly curved shape only (i.e.\ to pierce the drop). To get a feeling for how much the actual expansion rate of the drop depends on the laser focus we plot the maximum lateral expansion velocity $U_{x,{\rm max}}$ (see the inset for an illustration) at the drop surface as a function of $\sigma$ in Fig.\ \ref{fig6}a. One sees that a more focussed pulse does not only lead to a larger deformation but also to a larger lateral expansion: the smaller $\sigma$, the larger $U_{x,{\rm max}}$. In Fig.\ \ref{fig6}b we show at which position along the drop surface (in terms of azimuthal coordinate $\theta$) this maximum expansion velocity is observed. For a focussed pulse it is observed closer to the laser axis ($\theta=0$), whereas for an unfocussed pulse it is closer to the poles of the drop ($\theta=\pm\pi/2$).

\begin{figure}\begin{center}
	\includegraphics[width=\textwidth]{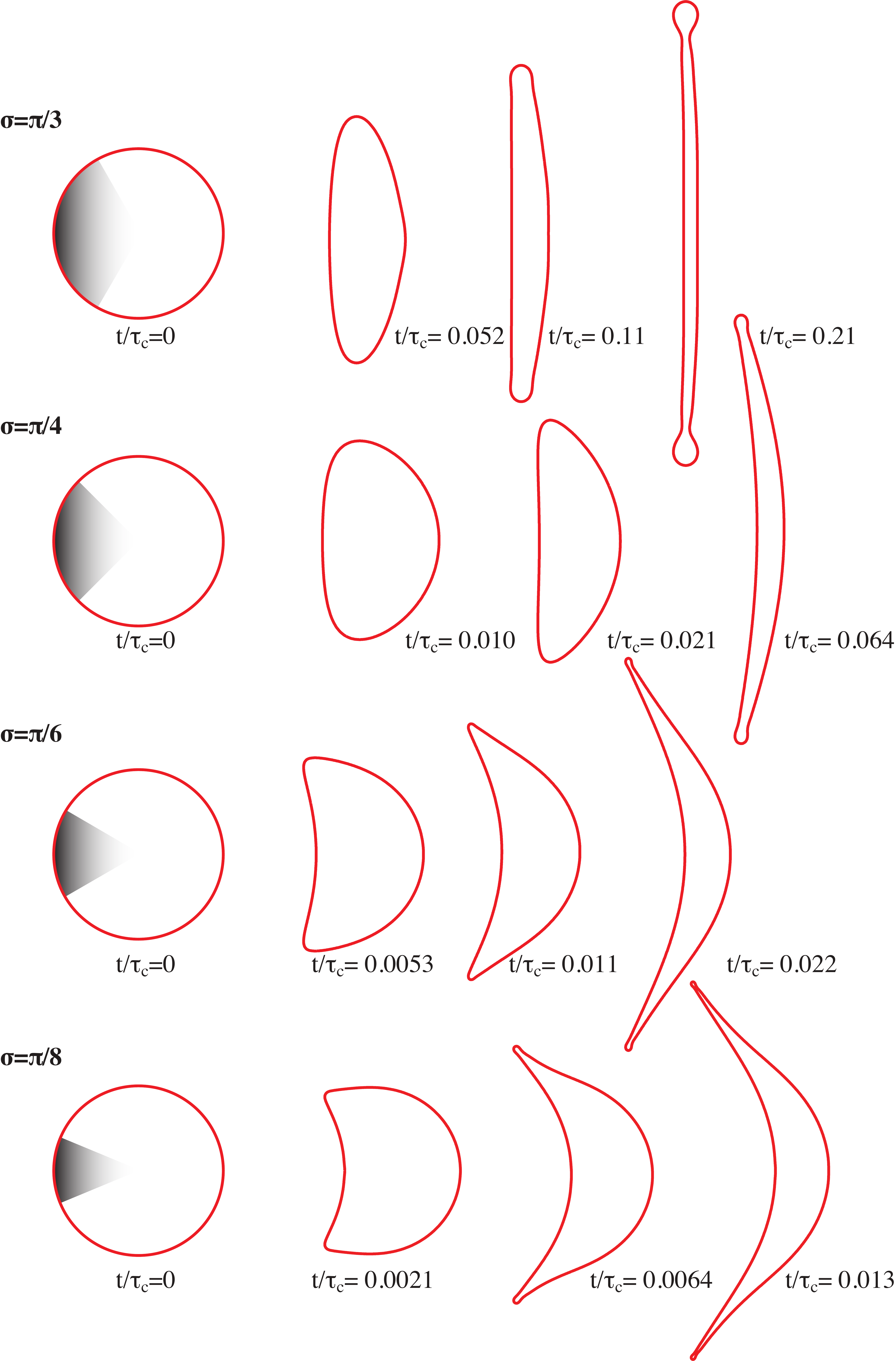}
	\caption{Sequences of drop contours obtained from the BI simulations illustrating the drop shape evolution for $\we=790$ and four different pulse widths $\sigma=\pi/3,~\pi/4,~\pi/6$ and $\pi/8$, from top to bottom. Clearly, a more focussed laser beam (smaller $\sigma$) leads to a larger expansion rate, a thinner sheet with a less uniform thickness and a more curved drop shape. Each sequence is sampled at different times to accommodate the different expansion rates.}\label{fig7}
\end{center}\end{figure}

\begin{figure}\begin{center}
	\includegraphics[width=\textwidth]{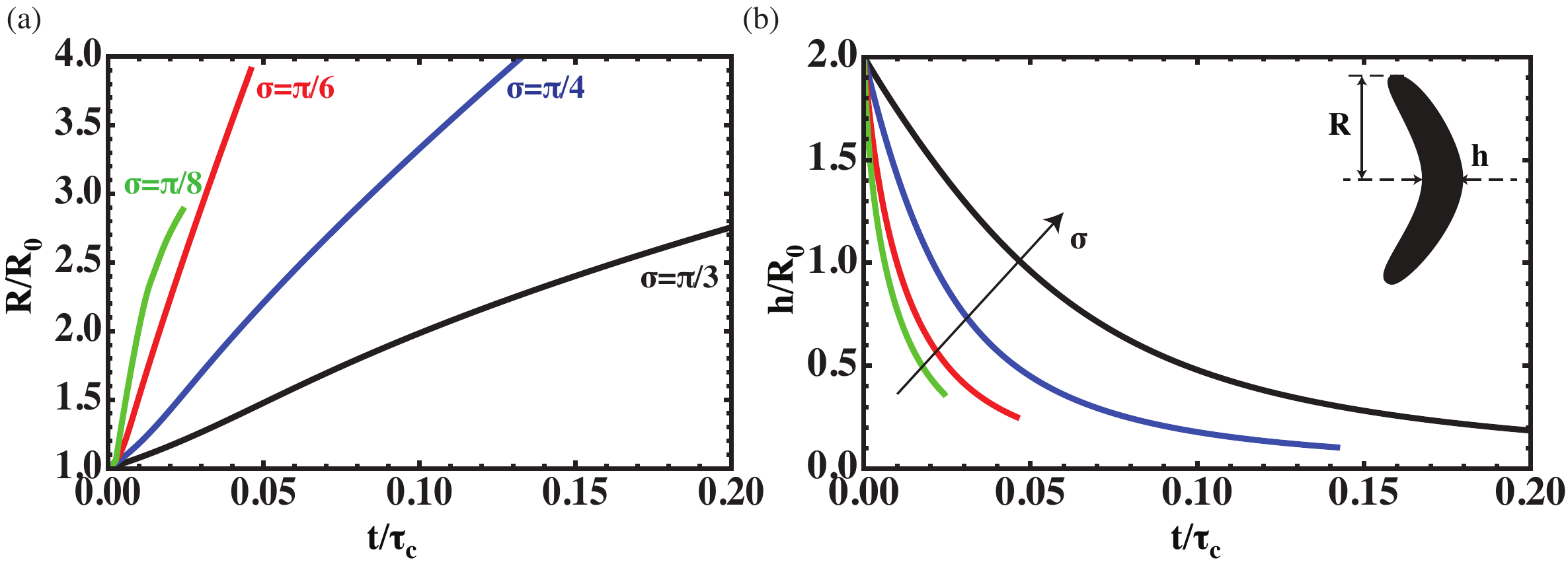}
	\caption{Sheet radius $R(t)$ (a) and thickness in the center $h(t)$ (b) extracted from the BI simulations shown in Fig.\ \ref{fig7} ($\we=790$ and $\sigma=\pi/8$, $\pi/6$, $\pi/4$ and $\pi/3$).}\label{fig8}
\end{center}\end{figure}

	The faster initial expansion rate for a focussed pressure pulse is confirmed by the simulations. Figure \ref{fig7} shows drop contours from the BI simulations for four different pulse widths, from which we derive the (projected) sheet radius $R$ and thickness $h$ (measured at the centre of the drop, see inset Fig.\ \ref{fig8}b). Indeed, in Fig. \ref{fig8} we observe that a smaller $\sigma$, i.e.\ a more focussed pulse, corresponds to a faster lateral expansion and a faster decrease in the sheet thickness. We therefore conclude that in order to get a maximally expanded sheet with a minimal thickness one needs to focus the laser pulse to a tight spot (spot size $\ll R_0$), with again the consequence that this maximally expanded sheet is strongly curved and has a non-uniform thickness (bottom panel Fig.~\ref{fig7}).

\subsection{A perfectly focussed laser pulse: the limit $\sigma\to 0$}\label{delta}

\begin{figure}\begin{center}
	\includegraphics[width=.3\textwidth]{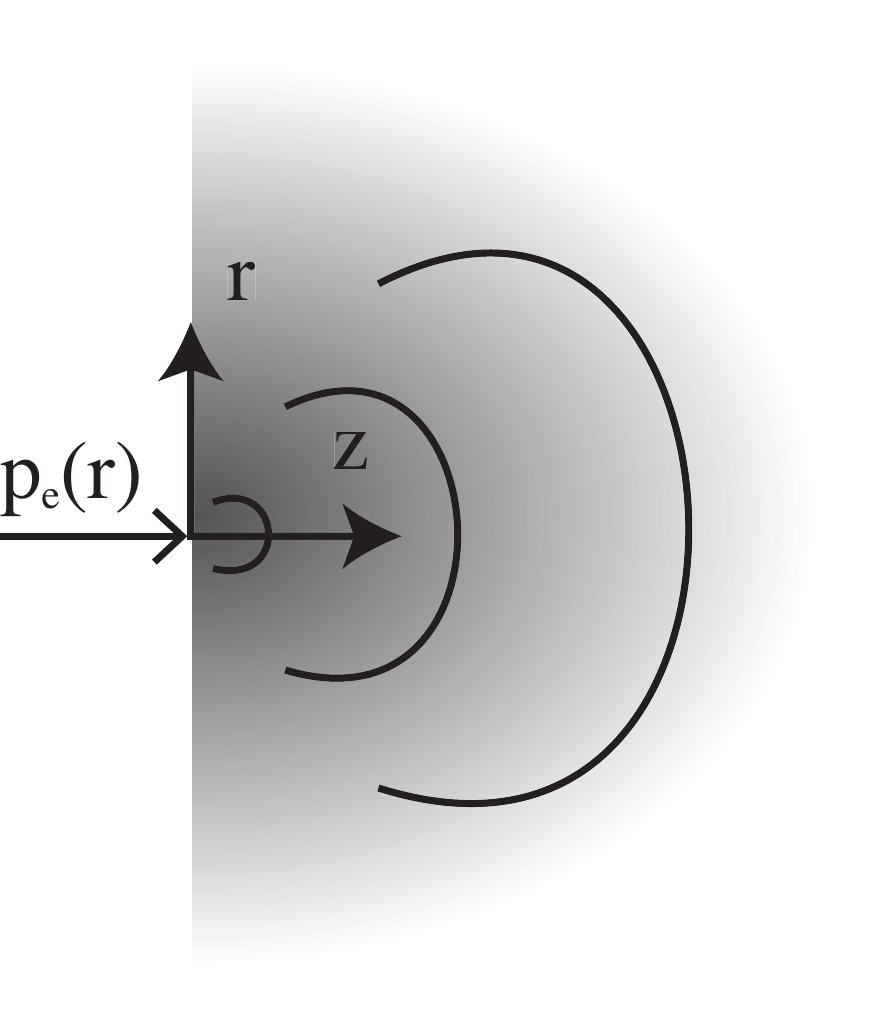}
	\caption{Situation for a tightly focussed laser beam. The drop surface curvature becomes irrelevant and the pressure pulse $p_{\rm e}$ comes down to a Dirac-delta distribution (\ref{bcdelta}) applying at the surface of an infinite half-space with cylindrical coordinates $r,z$.}\label{fig9}
\end{center}\end{figure}

	In the limit when the size of the laser pulse becomes negligibly small with respect to the drop size ($\sigma\to 0$) the pressure pulse on the drop surface approaches 
\begin{equation}
	f(\theta)\to \delta\left(\theta\right),\label{bcdelta}
\end{equation}
where $\delta$ stands for the Dirac-delta distribution and the series (\ref{sol1}) diverges. The exact solution to (\ref{ns2}, \ref{bcdelta}) can however be obtained from a different approach. For $\sigma \ll 1$ the curvature of the drop surface is no longer relevant and one recovers the response of an infinite half-space to a Dirac-delta pulse. To model this situation we adopt a cylindrical coordinate system ($r,z$), with the positive $z$-coordinate pointing into the liquid and $z=0$ corresponding to the liquid-air interface, see Fig.\ \ref{fig9}. 

	The boundary conditions for the Laplace equation (\ref{ns2}) in a half-space now read
\begin{eqnarray}
	& &p(r,z)\to 0 \quad \mathrm{for} \quad r,z\to\infty,\label{bc2}\\
	& &p(r,0)=\frac{\delta(r)}{2\pi r}.\label{bc3}
\end{eqnarray}
Hence, the pressure diverges at the origin, but the total force applied to the drop remains finite and equal to unity. The solution to the Laplace equation (\ref{ns2}) with boundary conditions (\ref{bc2},\ref{bc3}) is obtained by taking the Hankel transform of (\ref{ns2}) in $r$ \citep[\S 6.7]{Prosperetti:2011} from which we find, using (\ref{bc3})
\begin{equation}
	p(r,z)=\int_0^\infty s p(s,0)\mathrm{d}s\int_0^\infty k J_0 (kr) J_0(ks)e^{-kz}\mathrm{d}k=
\frac{1}{2\pi z^2\left(1+\left(r/z\right)^2\right)^{3/2}}.\label{soldel}
\end{equation}
The velocity field is then obtained from (\ref{velfield}):
\begin{eqnarray}
	u_r(r,z)&=&\frac{3rz}{2\pi \left(r^2+z^2\right)^{5/2}},\\
	u_z(r,z)&=&-\frac{r^2-2z^2}{2\pi \left(r^2+z^2\right)^{5/2}},
\end{eqnarray}
and diverges as $\epsilon^{-3}$ for $\epsilon=r, z \to 0$.  As a consequence, the total kinetic energy contained in the half-space is non-integrable. We therefore calculate the kinetic energy excluding a region of size $\epsilon$ around the origin 
\begin{equation}
	E_{\rm k}=\lim_{\epsilon\to 0} \pi\int_0^\infty \int_\epsilon^\infty \left(u_r^2+u_z^2\right) r \mathrm{d}r\mathrm{d}z=\lim_{\epsilon\to 0} \frac{3}{128\epsilon^3}. 
\end{equation}
Hence, the total kinetic energy diverges as $\epsilon^{-3}$ for $\epsilon\to 0$ and is contained in a tiny volume of size $\epsilon^3$, which is small compared to the drop size. In practice, for a Gaussian  of finite width (\ref{bc}) one can interpret $\epsilon=\sigma$ in the limit $\sigma\to 0$, and hence the drop kinetic energy diverges as $\sigma^{-3}$. We verified that the total kinetic energy obtained from the series solution (\ref{sol1}) indeed exhibits the same divergence. Since the translation kinetic energy of the drop is constant $E_{\rm k,d}/E_{\rm k}\to 1$ as $\sigma\to 0$, as was already observed in Fig.\ \ref{fig5}.

	The same conclusion can be reached from a simple scaling argument. We apply a finite force $F=\int p\mathrm{d}A$ to the drop.  In the limit $\sigma\to 0$ the characteristic area on which this force acts scales as $\sigma^2$, such that the local pressure $p\sim F/\sigma^2$ diverges. From momentum conservation the velocity field inside the drop scales as $u\sim F \tau/\rho \sigma^3$, such that the kinetic energy, which is used to deform a volume of size $\sigma^3$, scales as 
\begin{equation}
	E_{\rm k}\sim \rho u^2\sigma^3=F^2\tau^2/\rho \sigma^3,
\end{equation}
and hence diverges as $\sigma^{-3}$ for $\sigma\to 0$, while the ratio $E_{\rm k,d}/E_{\rm k}$ remains finite and approaches one.

\subsection{A one-sided uniform laser pulse yields a flat drop}\label{flat}

	A flat, symmetric drop shape can obviously be obtained by impacting the drop symmetrically with two laser beams. We will however see now that a flat shape can also be obtained with a uniform (or flat-top) laser-beam profile impacting the drop from one side only. 

\begin{figure}\begin{center}
	\includegraphics[width=\linewidth]{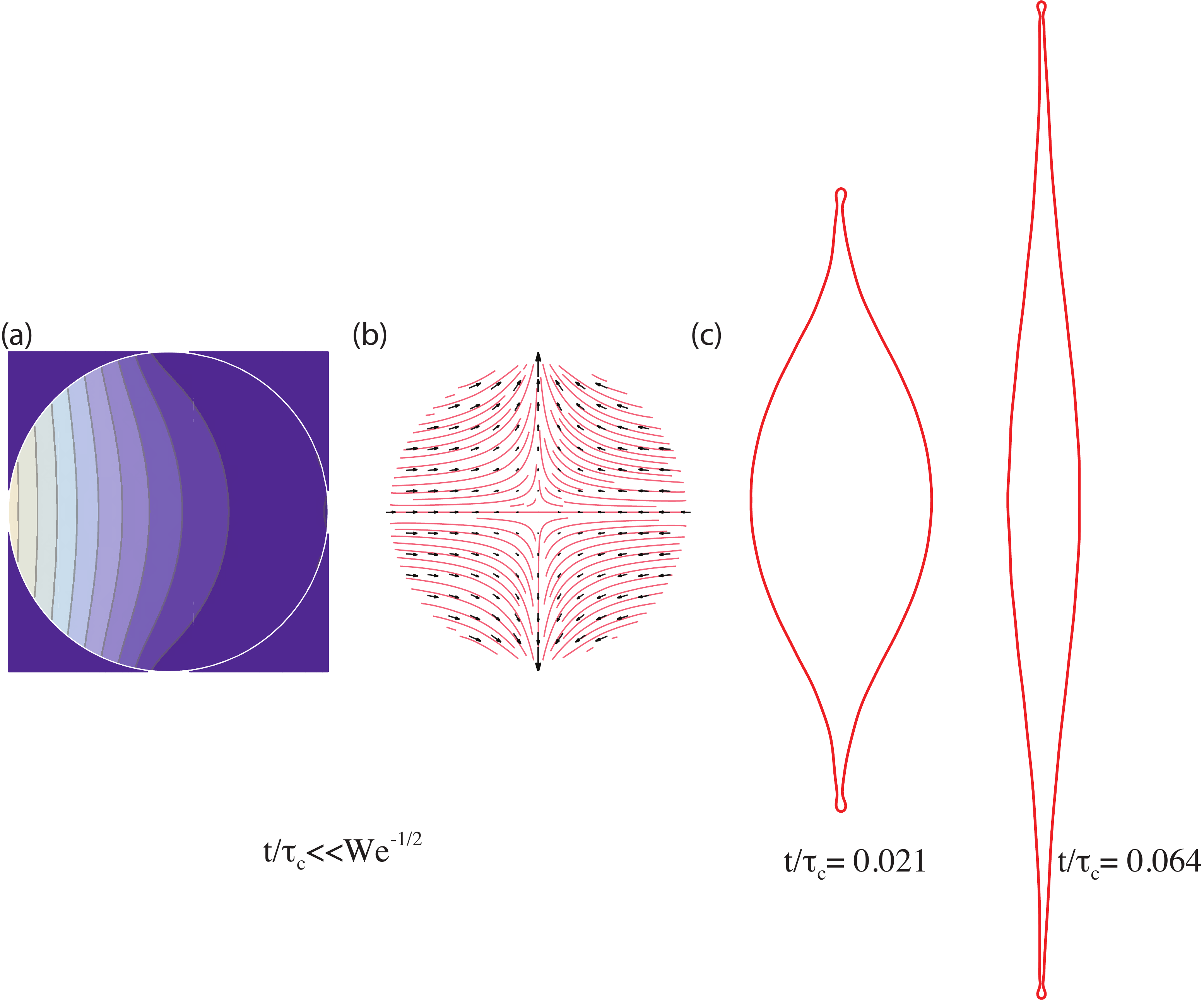}
	\caption{Symmetric deformation obtained for a uniform laser profile (cosine-shaped pressure pulse (\ref{coskick})). (a) Iso-pressure lines. (b) Streamlines of the velocity field in the co-moving frame. (c) Sequence of drop contours from the BI simulations for $\we=790$ (same scale as in (a-b)). }\label{fig10}
\end{center}\end{figure}

	As discussed above, a uniform laser-beam profile results in a cosine-shaped pressure profile on the drop surface
\begin{equation}
	f(\theta)=\frac{3}{2\pi}\cos\theta H\left(\pi/2-\theta\right),\label{coskick}
\end{equation}
where the Heaviside function $H$ restricts the interaction to the illuminated side of the drop. The coefficients (\ref{coef1}) can be obtained exactly and read
\begin{equation}
	A_\ell=\frac{3(2\ell+1)}{4\pi}\int_0^{\pi/2} P_\ell(\cos\theta) \cos \theta \sin\theta\mathrm{d}\theta=\frac{3\left(1+2\ell\right)}{16\sqrt{\pi}\Gamma(3/2-\ell/2)\Gamma(2+\ell/2)},
\end{equation}
from which we find
\begin{equation}
	p(r,\theta)=Ur\cos\theta +\frac{3}{16\sqrt{\pi}}\sum_{n=0}^\infty 
	\frac{1+4n}{\Gamma(3/2-n)\Gamma(2+n)}r^{2n}P_{2n}(\cos\theta),\label{flatpres}
\end{equation}
which involves only the even Legendre polynomials. The series (\ref{flatpres}) converges. However, despite the fact that the pressure field itself is continuous, its first derivative with respect to $\theta$ and hence the velocity $u_\theta$ exhibit a discontinuity in $\theta=\pi/2$ caused by the restriction of the pressure boundary-condition (\ref{coskick}) to the front of the drop. The resulting pressure (\ref{flatpres}) and velocity (\ref{velfield}) fields are shown in Fig.~\ref{fig10}. We use the velocity field to obtain the energy ratio
\begin{equation}
	\frac{E_{\rm k,d}}{E_{\rm k}}\approx 0.35.
\end{equation}
Notice that this energy ratio can also be reached with a Gaussian pressure pulse with $\sigma\approx 0.73$ (see Fig.\ \ref{fig5}).

	The velocity field in the co-moving frame shown in Fig.\ \ref{fig10}b displays a striking feature: it is symmetric not only around the horizontal axis (owing to the axi-symmetry of the pressure pulse), but also around the vertical axis. This means that the drop eventually deforms into a perfectly flat, symmetric shape even though the laser impact is only one-sided, as the BI results in Fig.~\ref{fig10}c confirm. One can understand this symmetry in the velocity field by inspecting the expression for the radial velocity after subtraction of the centre-of-mass velocity $U$
\begin{equation}
	u_r(r,\theta)=-\frac{3}{16\sqrt{\pi}}\sum_{n=1}^\infty \frac{2n(1+4n)}{8\Gamma(3/2-n)\Gamma(2+n)}r^{2n-1}P_{2n}(\cos\theta).\label{symvel}
\end{equation}
Realizing that $P_{2n}(x)=P_{2n}(-x)$, one sees that this velocity field is indeed symmetric. More heuristically one may note that a cosine pressure pulse accommodates the drop shape: both the pressure and the local thickness of the drop are proportional to $\cos\theta$ from which each slice of the drop acquires the same axial velocity. It can actually be proven that a cosine pressure pulse is the only one-sided profile that results in a symmetric (flat) drop, which we do in Appendix \ref{app}.

\section{Late time dynamics: the thin sheet limit}\label{latetimes}

 Up to now we have been concerned with the early time $t\sim \tau_{\rm e}$ of the dynamics, when the drop is still spherical and the influence of surface tension is negligible. The BI simulations allowed us to extend the description to later times, close to the maximal extension of the drop. We now consider the late-time regime $t\sim \tau_{\rm c}$ when the drop expands into a thin sheet and subsequently recedes.

\subsection{Problem formulation \& solution}
\begin{figure}\begin{center}
	\includegraphics[width=3 cm]{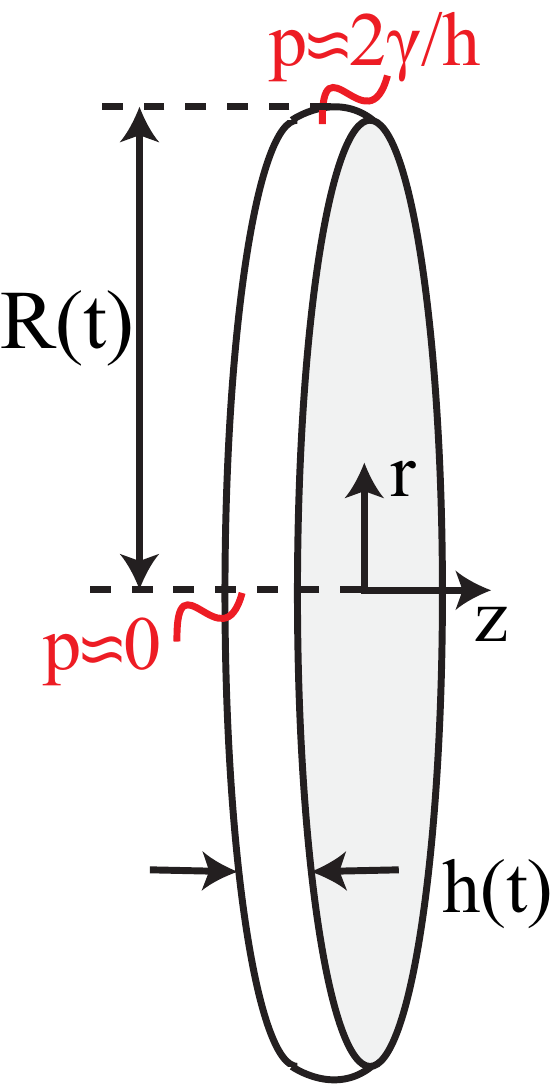}
	\caption{Sketch of an axisymmetric flat thin sheet with time-dependent thickness $h(t)$ and radius $R(t)$. The surface tension generates a typical Laplace pressure difference $2\gamma/h$ between the edge and the centre of the sheet, which drives the recession. The cylindrical coordinate system $(r,z)$ is indicated.}\label{fig11}
\end{center}\end{figure}

	We follow the same approach as \citet{Villermaux:2009} have used to describe the surface-tension limited expansion of a rain drop due to aerodynamics effects. We thus describe the dynamics of a flat, thin sheet in a frame co-moving with the centre-of-mass velocity, see Fig.\ \ref{fig11}. The sheet has a time-dependent, uniform thickness $h(t)\ll R(t)$ (with $R(t)$ the time-dependent drop radius) and a given initial kinetic energy, which is precisely that determined in the early-time model. This follows from the inviscid flow considered here: the kinetic energy is conserved as long as surface tension does not influence the drop deformation. Since we are typically interested in large Weber numbers, surface tension effects are negligible during the early stage of expansion ($t\sim \tau_{\rm i}\ll \tau_{\rm c}$). Furthermore, when the drop deforms into an essentially flat sheet, such as that shown in Fig.~\ref{fig4}a, all the kinetic energy of deformation is used to expand the drop laterally. One can therefore use the kinetic energy obtained from the early-time model as an initial condition for the thin-sheet model. 
 
	We adopt a cylindrical coordinate system $(r,z)$, with $r$ the lateral direction (in which the sheet expands) and $z$ the direction normal to the sheet surface (see Fig.~\ref{fig11}).
The sheet dynamics is prescribed by the axisymmetric Euler equation. In the thin-sheet approximation ($h\ll R$) the lateral flow $u(r,t)$ in the sheet is governed by
\begin{eqnarray}
	& &\frac{\partial u}{\partial t}+u_r\frac{\partial u_r}{ \partial r}=-\frac{\partial p}{\partial r},\label{mom}\\
	& &r\frac{\partial h}{\partial t}+\frac{\partial}{\partial r}\left( r uh \right)=0,\label{mass}
\end{eqnarray}
where all lengths are non-dimensionalised by $R_0$, all times by $\tau_{\rm c}$ and the pressure by $\gamma/R_0$. It follows from global mass conservation that $h=4/3R^{-2}$ so that using (\ref{mass}) we find
\begin{equation}
	u(r,t)=r\frac{\dot{R}}{R}.
\end{equation}
Integration of (\ref{mom}) between $r=0$ and $r=R(t)$ gives \citep{Villermaux:2009}
\begin{equation}
	R\ddot{R}=-2\left[p(R)-p(0)\right].\label{pancake1}
\end{equation}
For $r\ll R(t)$ the interface curvature is close to zero, whereas for $r=R$ it is approximately $2/h(t)$, such that (\ref{pancake1}) reduces to
\begin{equation}
	R\ddot{R}=-\frac{4}{h}=-3R^2.\label{pancake2}
\end{equation}
The solution reads $R(t)=a \cos\sqrt{3}t+b\sin \sqrt{3}t$, with constants $a$ and $b$ to be determined from the initial conditions. The initial radius $R(0)=1$ sets $a=1$. To derive the initial rate of expansion $\sqrt{3}b$ we use the fact that at $t=0$ the deformation kinetic energy of the sheet 
\begin{equation}
	E_{\rm k,d}^s=\frac{1}{2}\int_0^{R(t)}2\pi u^2h r\mathrm{d}r=\frac{1}{3}\pi\dot{R}^2\label{esheet1}
\end{equation}
has to match the deformation kinetic energy of the drop obtained from the early-time model. In terms of the early-time kinetic energy partition $E_{\rm k,d}/E_{\rm k,cm}$ we obtain
\begin{equation}
	E_{\rm k,d}^s(t=0)=\pi b^2=\frac{2}{3}\pi \left( \frac{E_{\rm k,d}}{E_{\rm k,cm}}\right)\we.\label{enexp}
\end{equation}
Eliminating $b$ from (\ref{enexp}) we find the solution
\begin{equation}
	R(t)=\cos\sqrt{3}t+\left(\frac{2}{3}\right)^{1/2}\left(\frac{E_{\rm k,d}}{E_{\rm k,cm}}\right)^{1/2}\we^{1/2}\sin \sqrt{3}t. \label{pcdyn}
\end{equation}
This square-root dependence of the sheet radius on the Weber number is well-known for drop impact on solids in absence of friction \citep{Villermaux:2011}. As \citet{Klein:2015} already showed, it is also in good agreement with experimental observations for a drop impacted by a laser. 

	It is important to realize that the expanding thin sheet described here is actually subjected to hydrodynamic instabilities that may eventually cause the sheet to fragment, as Fig.\ \ref{fig1}b-d clearly shows. First, the rapid acceleration of the drop on $\tau_{\rm e}$ may trigger a destabilization in the sense of Rayleigh-Taylor, which could puncture the sheet, similar to what has been observed by \citet{Bremond:2005} for sheets subjected to shock waves. Second, the rim formed at the edge of the receding sheet may develop both Rayleigh-Taylor and Rayleigh-Plateau instabilities, as it is observed for a drop impacting a pillar \citep{Villermaux:2009}. A description of these instabilities is however beyond the scope of the present paper and is left for future work.

\subsection{Comparison to BI and experiments}

	To compare the thin-sheet model (\ref{pcdyn}) to the experimental and BI results presented by \citet{Klein:2015} we determine the energy partition resulting from the experimental beam profile. \citet{Klein:2015} showed that the latter is well described by a Gaussian curve (\ref{bc}) of width $\sigma=\pi/6$. In the same study, this Gaussian pressure profile was already successfully used in BI simulations to calculate the lateral drop expansion, which suggests that irregularities in the beam profile have a negligible influence on the drop expansion. Here we use the same pressure profile in the early-time model to determine the kinetic energy partition 
\begin{equation}
	\frac{E_{\rm k,d}}{E_{\rm k,cm}}=1.8,\label{esheet}
\end{equation}
or, in terms of the energy ratio depicted in Fig.~\ref{fig5}, $E_{\rm k,d}/E_{\rm k}=0.64$.
\begin{figure}\begin{center}
	\includegraphics[width=10 cm]{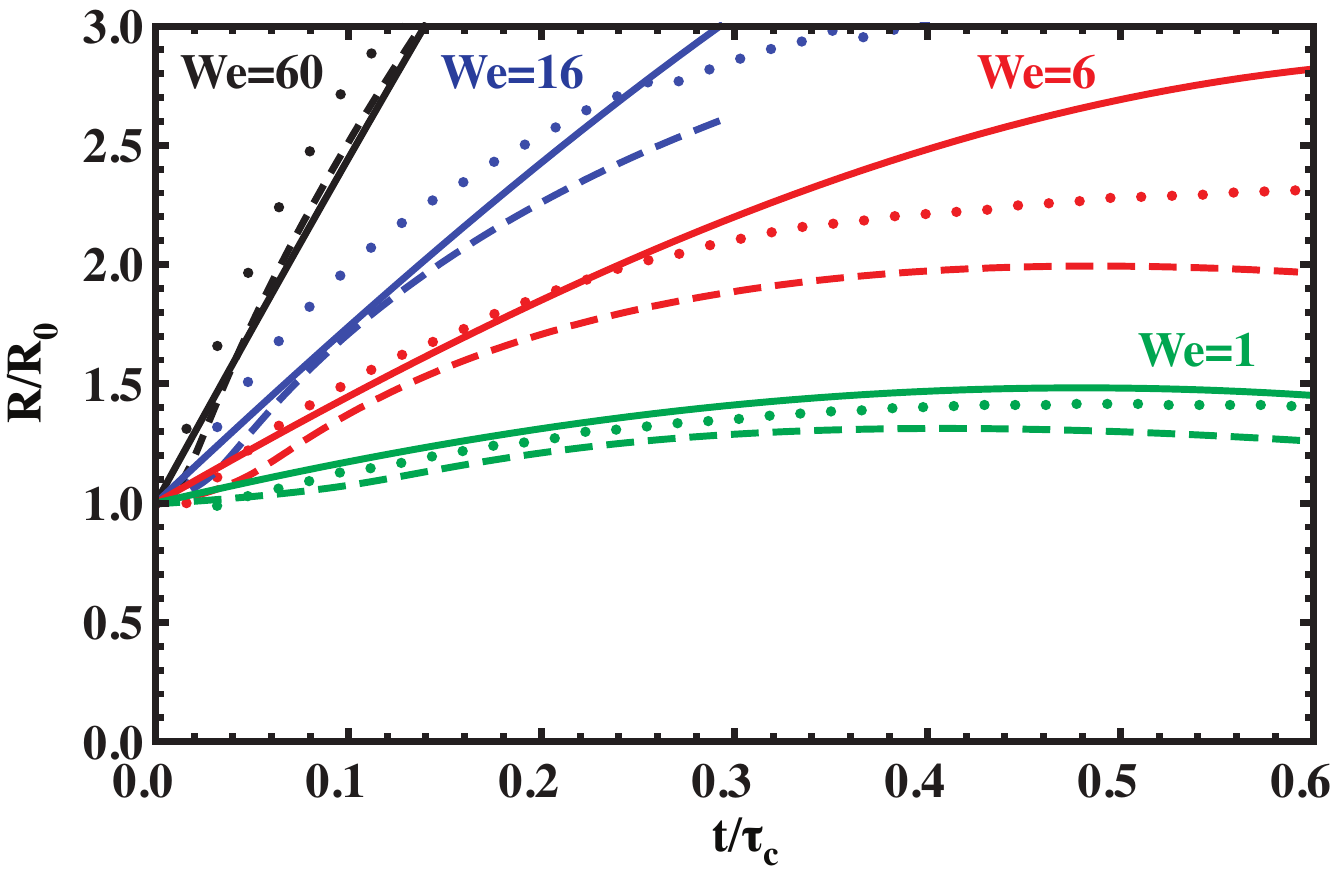}
	\caption{Time evolution of the sheet radius $R$ for $\we=60$ (black), $\we=16$ (blue), $\we=6$ (red) and $\we=1$ (green). The dashed lines represent the BI simulations for $\sigma=\pi/6$ and the dots are the experimental data, both reproduced from \citet{Klein:2015}. The solid lines shows the theoretical prediction based on the thin sheet model (\ref{pcdyn}) with no adjustable parameter: the initial radius is set to unity and the initial kinetic energy partition is taken from the inertial model for a pulse width $\sigma=\pi/6$ that matches the experimental beam profile. Experimental and BI data are shown until the fragmentation starts to influence the sheet radius or the sheet becomes locally too thin to be resolved accurately in the numerics, respectively.}\label{fig12}
\end{center}\end{figure}

	Figure \ref{fig12} compares the thin-sheet model (\ref{pcdyn}) with experimental and BI results. The thin-sheet model assumes a flat drop, whereas in the experiments and BI simulations the drop is curved (for a pulse width $\sigma=\pi/6$, see Fig.~\ref{fig7}).
	For the comparison we therefore use the projected radius as defined in the inset in Fig.\ \ref{fig8}b. The thin-sheet model provides an accurate prediction of the initial expansion speed and temporal evolution of the sheet for all Weber numbers without any adjustable parameter. For $\we=60$ the BI and theoretical model almost completely overlap and are very close to the experimental data. Not surprisingly, at smaller Weber numbers, when the drop does not expand into a thin sheet with $h\ll R$ anymore, both models show the same initial expansion rate but start to deviate at later times. In particular, for $\we\sim 1$  the drop only oscillates around its spherical shape (see Fig.~1 in \citet{Klein:2015}) and the thin-sheet approximation obviously breaks down.

	The good agreement between the thin-sheet model and the experiments and BI simulations suggests that, although the initial expansion rate is very sensitive to the beam width (see Figures \ref{fig5} and \ref{fig6}) moderate curvatures and thickness variations in the sheet have little influence on the actual global expansion of the drop. One should however bear in mind that these moderate non-uniformities might have important consequences for the eventual sheet puncture and fragmentation.

%At large Weber numbers ($\we=16, 60), a strong overshoot of the thin sheet model with respect to, in particular, the experimental data is observed. An explanation for this is the fragmentation at the edge of the sheet that happens in experiment at these larger Weber numbers corrugates the measurement of the actual sheet radius. These drop ejection events from the edge occur already early in time ($t/\tau_{\rm c}\sim 0.1-0.2$, see \citet{Klein:2015}). These fragmentation events are neither included in the thin sheet model, nor described by the BI simulations, which are axisymmetric and hence do not allow for instabilities to develop along the sheet edge. 

\section{Conclusion}

	The interaction of a laser pulse with an absorbing liquid drop can successfully be modelled by applying a recoil-pressure pulse to the drop surface. The relation between the total impulse of this recoil pressure and the laser-pulse energy is found from scaling arguments, whereas the profile of the pressure pulse can be considered, as a first approximation, to follow that of the drop surface illumination (i.e.~that of the laser-beam profile weighted by its local incidence on the drop surface). Once this relation is known, the hydrodynamic response of the drop to the laser impact (propulsion, expansion and recession, possibly leading to fragmentation)  is entirely captured from the drop response to the corresponding pressure pulse. This approach allows to study the response of the drop to laser pulses of different shapes and focus. 

	An analytical model for the impulsive acceleration when the drop is still spherical provides the early-time drop dynamics as a function of the laser-pulse shape: the kinetic-energy partition inside the drop is obtained, from which we derive the amount of deformation versus translation of the drop. This yields a first-order estimate of the drop shape evolution at later times by advecting the material points on the drop surface. We find that, for a given propulsion of the drop, a maximal expansion is obtained when the laser pulse is focussed into a tight spot, which results in a strongly curved sheet, while a flat symmetric sheet can only be obtained with a uniform laser-beam profile. 
	
	On the inertial and capillary timescales boundary integral simulations reveal the detail of the sheet thickness and curvature dependence on the pulse focus, until close to the maximal expansion (where the simulation breaks down). Assuming a flat drop, we derive an analytical thin-sheet model initialized with the expansion rate obtained from the early-time model. The thin-sheet model predicts the entire evolution of the sheet radius (expansion and recession) and shows a good agreement with both experimental and BI data, in particular for large Weber numbers.

	The drop deformation dynamics described by the models discussed here forms the starting point to study the subsequent drop fragmentation which is observed experimentally for high-energy laser pulses (i.e.\ drop expansion at large Weber number), see Fig.\ \ref{fig1}b-d in the present paper and also Fig.~1 in \citet{Klein:2015}. Understanding the mechanisms behind this fragmentation will be the subject of future work.

%	Last, it should also be realized that beside the essentially inviscid expansion observed in the experiments we discuss here, different situations are possible. The Reynolds number of an inviscid deformation is typically $R_{\rm max}^2/\nu\tau_{\rm c}$, where $\nu$ is the liquid viscosity, which equivalently writes $\we/\oh$ in terms of the Ohnesorge number $\oh = \sqrt{\nu^2/\gamma R_0/\rho} $. One precisely interested in large expansions up to the onset of fragmentation shall consider $\we \sim 10^2$, hence a typical Reynolds number $10^2/\oh\propto \sqrt{R_0}/\nu$. The latter remains fairly large (say $\gtrsim 10$) down to micrometer-sized water droplets, but easily becomes smaller than unity with moderately viscous liquids even for millimeter-sized drops. For such small or viscous drops, a highly damped dynamics has to be expected, which would also worth being studied.   
% open issue: understanding what sets pressure pulse duration $\tau_{\rm e}$, more insight in propulsion mechanism

\begin{acknowledgements}
	We are grateful to Andrea Prosperetti, Howard A.\ Stone, Leen van Wijngaarden and Andrei Yakunin for valuable discussions. 
This work is part of an Industrial Partnership Programme of the Foundation for Fundamental Research on Matter (FOM), which is financially supported by the Netherlands Organization for Scientific Research (NWO). This research programme is co-financed by ASML. % This research is financially supported by FOM, NWO and ASML
\end{acknowledgements}

\appendix
\section{Only a uniform laser beam profile results in a flat drop}\label{app}
	
	In \S \ref{flat} we found that a one-sided impact with a uniform laser beam (and hence a cosine-shaped pressure pulse) results in a flat symmetric drop. We demonstrate here that the uniform beam profile is in fact the only profile that gives rise to a flat drop.

	A requirement for symmetry is that after subtraction of the centre-of-mass speed the velocity field satisfies the property $u_r(r,\theta)=u_r(r,\pi-\theta)$. Inspecting e.g.\ (\ref{symvel}) we see that this requires the odd coefficients in (\ref{coef1}) to be equal to zero, except for $A_1=U$ to ensure the centre-of-mass speed. Since the velocity field (\ref{velfield}) is obtained from the pressure field by taking the gradient, the symmetry of the velocity field implies that the odd coefficients in the pressure field (\ref{sol1}) should also be equal to zero (again except for $A_1$). Hence, the pressure pulse $f$ (\ref{presbc}) needs to satisfy 
\begin{equation}\label{aprop}
\begin{array}{rl}
	&A_1=\frac{3}{2}\int_{-1}^1 f(x)x\mathrm{d}x=U,\\
	&A_{2n+1}=\frac{4n+3}{2}\int_{-1}^1f(x)P_{2n+1}(x)\mathrm{d}x=0~\mathrm{for}~n>0.
\end{array}
\end{equation}
When the drop is hit by a laser pulse the recoil pressure is only exerted from one side:
\begin{equation}
	f(x)=g(x)H(1-x).\label{oneside}
\end{equation}
Hence, we need to find the functional form of $g$ such that $f$ satisfies (\ref{aprop}). To this end, we express $g$ into the Legendre series
\begin{equation}
	g(x)=\sum_{m=0}^\infty d_m P_m(x).\label{g}
\end{equation}
Substituting (\ref{oneside},\ref{g}) into (\ref{aprop}) and evaluating the coefficients $A_n$ we obtain 
\begin{equation}
	A_{2n+1}=\frac{4n+3}{2}\sum_{m=0}^\infty d_m\int_0^1P_m(x)P_{2n+1}(x)\mathrm{d}x, ~\mathrm{for}~n=0,1,2 \dots,
\end{equation}
where the integral now runs from zero to one.
Using the property that
\begin{equation}
	\int_0^1P_m(x)P_n(x)\mathrm{d}x=\left\{ \begin{array}{ll}
      \frac{1}{2m+1} & \mbox{if $m=n$,}\\
       0 & \mbox{if $m\neq n$, $m,n$ both even or odd,}\\
          h_{m,n} & \mbox{if $m$ even, $n$ odd,}\\
        h_{n,m} & \mbox{if $m$ odd, $n$ even},
        \end{array} \right.
\end{equation}
with $h_{m,n}=\frac{(-1)^{(m+n+1)/2}m!n!}{2^{m+n-1}(m-n)(m+n+1)\left[(\frac{1}{2}m)!\right]^2\left\{[\frac{1}{2}(n-1)]!\right\}^2} $ \cite[p. 173]{Byerly} we find that 
%\begin{equation}
%\begin{array}{rl}
%&A_1=\frac{3}{2}\left(\frac{1}{3}d_1+\sum_{m=0}^\infty d_{2m} h_{2m,1}\right),\\
%&A_{2n+1}=\frac{4n+3}{2}\left(\frac{1}{4n+3}d_{2n+1}+\sum_{m=0}^\infty d_{2m}h_{2m,2n+1}\right)~\forall~n>0.
%\end{array}
%\end{equation}
\begin{equation}
	A_{2n+1}=\frac{4n+3}{2}\left(\frac{1}{4n+3}d_{2n+1}+\sum_{m=0}^\infty d_{2m}h_{2m,2n+1}\right),~\mathrm{for}~n=0,1,2 \dots. \label{coefhalf}
\end{equation}
In order to satisfy (\ref{aprop}) we need $A_{2n+1}=0 ~\forall~ n>0$. From (\ref{coefhalf}) we observe that this requirement is satisfied for all $n$ simultaneously only when $d_{m}=0 ~\mathrm{for}~m \neq 1$, i.e.
\begin{equation}
	g(x)=d_1P_1(x)=d_1\cos(x).
\end{equation}
%We have verified numerically up to $n=1000$ that the system of equations $A_{2n+1}=0$ indeed has as only solution $d_m=0$ for $m\neq 1$.
This implies that the only way to form a flat, symmetric drop with a one-sided impact is to illuminate the drop uniformly, i.e. with a uniform or strongly defocussed Gaussian laser-beam profile.

\section{Method to determine the kinetic-energy partition in experiments}\label{app2}

	In Fig.\ \ref{fig5} we showed the analytically obtained kinetic-energy partition in the drop as a function of the pulse width $\sigma$. For comparison, we also plotted the data points corresponding to the experiments shown in Fig.\ \ref{fig1}. Below we outline how these experimental estimates are obtained.
 
	In case the drop expands into a flat, thin sheet all deformation kinetic energy is used for lateral expansion and we find, using (\ref{esheet1}), the kinetic-energy partition 
\begin{equation}
	\frac{E_{\rm k,d}^s}{E_{\rm k,cm}}=\frac{1}{2}\frac{\dot{R}^2}{U^2}
\end{equation}
and hence
\begin{equation}
	\frac{E_{\rm k,d}^s}{E_{k}}=\frac{\dot{R}^2}{\dot{R}^2+2U^2}.\label{epartsheet}
\end{equation}
The above expression is exact in case the drop expands into a flat sheet, hence for a uniform laser-beam profile. However, for the experimental data points shown in Fig.~\ref{fig5} we also used ({\ref{epartsheet}) to estimate the energy partition for more focussed beam profiles.
To obtain this estimate we had to extract the lateral expansion rate and the centre-of-mass speed of the drop for the different cases shown in Fig.~\ref{fig1} from simultaneous high-speed front- and side-view recordings of the drop shape evolution (for details on the experimental set-up the reader is referred to \citet{Klein:2015}). 

We determined the initial expansion rate $\dot{R}$ based on the first three images available in the front-view recordings by fitting ellipses to the drop shape at each instant, as explained in Fig.~\ref{fig:dotR}. 
The selected frame rate of 10\,000 frames per second ensures a sufficiently rapid sampling of the expansion such that the first three data points are well described by a linear fit (Fig.~\ref{fig:dotR}b). Difficulties in the determination of the actual equivalent drop radius $R$ arise when ligaments formed by the fragmentation of the drop corrupt the view (see e.g.~Fig.~\ref{fig1}b). In our analysis, we excluded these ligaments from the ellipse-fitting.
\begin{figure}\centering
	\includegraphics[width=\linewidth]{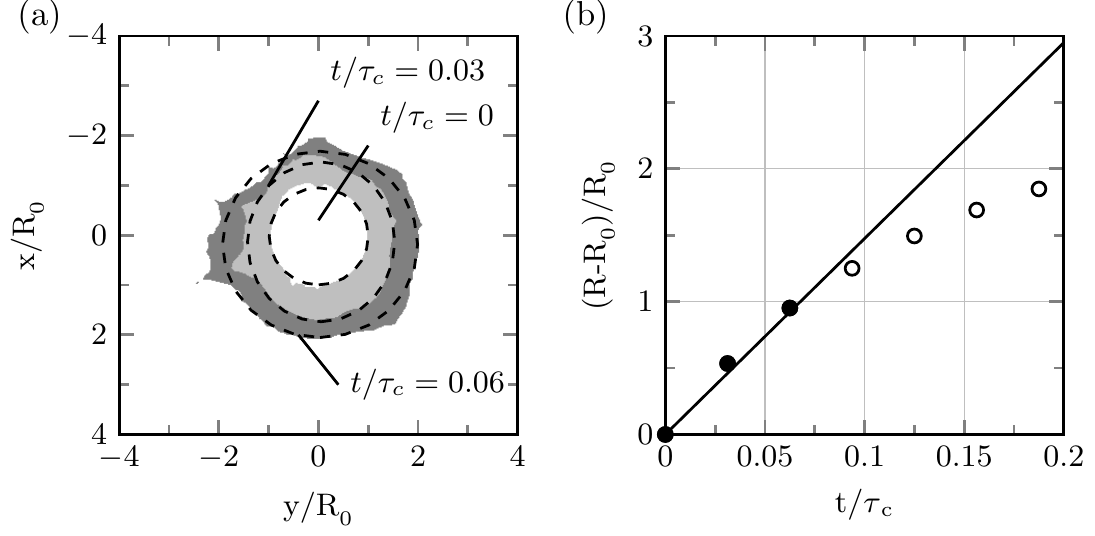}
	\caption{(a) Stacked front-view images (taken from the laser beam axis) of the expanding drop shown in Fig.~\ref{fig1}c. The dashed lines are the ellipses best fitting the contours, from which the equivalent radii $R$ are determined. (b) Relative expansion of the drop obtained from the views shown in (a) (\protect\includegraphics{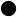}) and from later times (\protect \includegraphics{figure13_legend02.pdf}). The linear fit (solid line) to the first points yields a dimensional expansion speed $\dot{R} = 14.6\, R_0/\tau_{\rm c}$ = 4.2\,m/s.}
	\label{fig:dotR}
\end{figure}

	To determine the propulsion speed $U$ of the drop we used the side-view images (see Fig.~\ref{fig:U}). We assumed an axisymmetric drop shape and determined the center-of-mass position $z_{cm}$ for each frame of the high-speed recordings. 
After the initial acceleration of the drop on the timescale $\tau_{\rm e}$ the propulsion speed is constant and can hence be determined by a linear fit to the center-of-mass position. Since the side-view images are two-dimensional projections of the actual drop shape, they do not resolve the concavity of the drop.  This introduces an uncertainty in the determination of the center-of-mass position, in particular for the more focussed laser-beam profiles, where the drop evolves into a concave shape. We estimate the total error in $\dot{R}/U$ due to all the effects described above to be of the order of $20\%$.
% However, the error is minimized by choosing a sufficiently long tracking interval that covers time steps where the concave shape has not yet developed, i.e. $t/\tau_{\rm c} = 0$, or has already collapsed due to surface tension, i.e. $t/\tau_{\rm c} = 1$.% Therefore, any two consecutive images with a temporal delay $\Delta t \gg \tau_{\rm e}$ can be used to measure the propulsion speed by the change of the center-of-mass position $z_{cm}$ with the temporal delay between the images, i.e. $U = z_{cm}/\Delta t$.
\begin{figure}\centering
	\includegraphics[width=\linewidth]{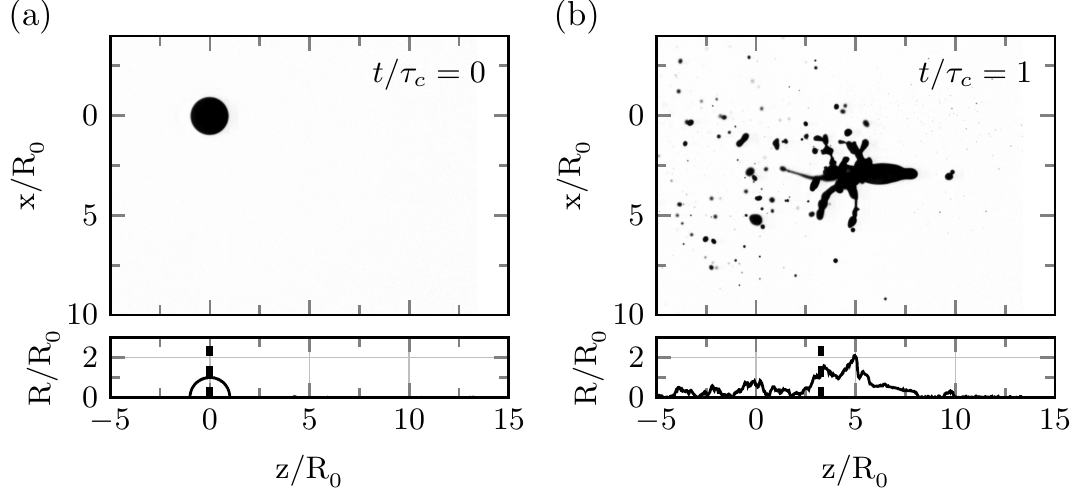}
	\caption{Two side-views (taken perpendicularly to the laser-beam direction) of the event shown in Fig.~\ref{fig1}d at (a) $t/\tau_{\rm c} = 0$ and (b) $t/\tau_{\rm c}=1$. The top frames show the actual shadowgraphs and the bottom ones show an axisymmetric profile that was obtained by summing up pixel values in $x$ direction (i.e.~collapsing all pixel values to the axis). The dashed line indicates the center-of-mass axial position $z_{cm} = \int z R^2 \mathrm{d}z\, /\, \int R^2 \mathrm{d}z$, which assumes axisymmetry of the profile.% The displacement leads to a corresponding speed of $U = 3.3\,R_0/\tau_{\rm c}$. % $z_{cm} = \int_{-5}^{15} z R^2 \mathrm{d}z\, /\, \int_{-5}^{15} R^2 \mathrm{d}z$ $z_{cm} = \frac{\int z R^2 \mathrm{d}z}{\int R^2 \mathrm{d}z}$
	% Measurement of the propulsion speed $U$ illustrated for the experiment shown in Fig.~\ref{fig1}d. 
	% The drop position along the laser beam propagation direction $z$ is based on the center-of-mass of the drop in shadowgraph images at two points in time $t/\tau_{\rm c} = 0$ and 1, figure (a) and (b) respectively. 
	% The center-of-mass is obtained from an axi-symmetric representation by adding pixel values along the radial direction $x$ pix$U = 3.3\,R_0/\tau_{\rm c}$ for THIRD FIGURE
	}\label{fig:U}
\end{figure}

	Finally, for each experiment the corresponding laser-pulse width $\sigma$ was determined by fitting our experimental laser-beam profiles with a Gaussian curve; see \citet{Klein:2015} for details. Errors may arise from deviations of the beam profile from a perfect Gaussian, shot-to-shot variations in the laser-beam profile and uncertainty in the drop position within the laser beam, in particular for the unfocussed beams where the drop is still hit by the laser even if it is positioned slightly off-centre. For the experiments shown in Fig. \ref{fig1} we estimate the uncertainty in $\sigma$ to vary from $\sim 15\%$ for the unfocussed case to $\sim 5\%$ for the most focussed case.

	To investigate the validity of our estimate for the energy partition (\ref{epartsheet}) we use the results from the BI simulations shown in Fig.\ \ref{fig7} and \ref{fig8}. In BI the centre-of-mass speed is known (and constant for each value of $\sigma$) and the pressure profile is exactly Gaussian with a known $\sigma$. Hence, in BI the uncertainties that appear in experiment are absent and the only approximation that remains is the use of (\ref{epartsheet}) as a measure for the energy partition. To find $\dot{R}$  in BI we determined the initial slopes of the curves in Fig.~\ref{fig8}a, similar to what has been done for the experimental data. The resulting estimate for the energy partition is in good quantitative agreement with its theoretical prediction; see the red squares in Fig.~\ref{fig5}. Small deviations ($<15\%$) are observed for the most focussed pulses ($\sigma=\pi/6$ and $\pi/8$), where the sheet is strongly curved and hence the approximation breaks down. Nevertheless, the quantitative agreement between theory and BI confirms that (\ref{epartsheet}) is indeed a reasonable estimate of the energy partition in the range of pulse widths studied here.

\end{document}